\documentclass[11pt]{article}
\pdfoutput=1 
\usepackage{comment}
\usepackage{jheppub}
\usepackage{dashrule}
\usepackage[T1]{fontenc} 
\usepackage{braket}
\usepackage{float}
\usepackage{amsmath,amssymb}
\usepackage{graphicx}
\usepackage{textcomp}
\usepackage{mathtools}
\usepackage{eucal}
\usepackage{mathrsfs}
\usepackage{tikz}
\usepackage{makecell}
\usepackage{ytableau}
\usepackage[bbgreekl]{mathbbol}
\usepackage{soul}

\newcommand{\vx}{\vec{x}}
\newcommand{\vxp}{\vec{x}\,{}'}
\newcommand{\rhop}{\rho'}
\newcommand{\etaret}{\eta_{\rm ret}}
\newcommand{\etaz}{\eta_z}
\newcommand{\fh}{\mathfrak{h}}
\newcommand{\Th}{\Theta}
\newcommand{\Boxf}{\Box}
\newcommand{\dd}{d}
\newcommand{\pd}{\partial}
\newcommand{\scri}{\mathscr I}
\def\be{\begin{equation}}
\def\ee{\end{equation}}
\def\bea{\begin{eqnarray}}
\def\eea{\end{eqnarray}}
\newcommand{\Dm}{D_{-}}

\newcommand{\hn}{\widehat{n}}
\newcommand{\hxp}{\widehat{x}\,{}'}
\newcommand{\chihat}{\widehat{\chi}}
\newcommand{\cS}{\mathcal{A}}
\newcommand{\fQ}{\mathcal{Q}}
\newcommand{\fP}{\mathcal{P}}
\newcommand{\fS}{\mathcal{S}}
\newcommand{\That}{\widehat{T}}

\newcommand{\Dp}{D_{+}}

\newcommand{\STF}{\text{STF}}

\newcommand{\Qrp}{Q^{(\rho+p)}}

\newcommand{\aR}{a_{\mathrm{ret}}}

\makeatletter
\g@addto@macro\bfseries{\boldmath}\g@addto@macro\bfseries{\boldmath}
\makeatother
\title{
Linearised gravitational radiation in de Sitter spacetime at all multipole orders
}
\author{Harsh$^a$,
Omkar Shetye$^{b}$,  Amitabh Virmani$^a$}
\emailAdd{harsh22@cmi.ac.in, omkar.sanjay.shetye@uj.edu.pl, avirmani@cmi.ac.in}
\affiliation[a]{Chennai Mathematical Institute, 
H1, SIPCOT IT Park, Siruseri, Kelambakkam 603103 India}
\affiliation[b]{ Instytut Fizyki Teoretycznej,
  Uniwersytet Jagiello\'nski,
  ul. \L{}ojasiewicza 11,
  30-348 Krak\'ow,
  Poland
}

\abstract{We derive the linearised gravitational waves generated by a localised matter source in the Poincar\'e patch of de~Sitter spacetime at arbitrary multipole order. The linearised de~Sitter problem is mapped exactly to an auxiliary flat-space problem in harmonic gauge, with a conserved source built from the physical stress tensor. This allows us to use the multipole expansion of Damour and Iyer verbatim, and yields the exterior metric at all orders in terms of two families of radiative multipole moments given as explicit integrals of the physical stress tensor. The de~Sitter tail is absorbed entirely into the auxiliary stress tensor, so that neither the moments nor the canonical exterior metric carry an integral over the past history of the source. 
As a check, we recover the known quadrupolar metric: in the quadrupolar truncation it is our canonical field up to
a residual gauge transformation, which we construct explicitly.}

\begin{document}
\maketitle

\section{Introduction}
Our Universe has a positive cosmological constant $\Lambda$. Gravitational radiation is
characterised by the field far from its source, and at large enough distances the
cosmological term dominates the field of any isolated system, however small $\Lambda$ is.
The asymptotic geometry is thus set by $\Lambda$, and for $\Lambda>0$ the spacetime of an
isolated system is asymptotically de~Sitter rather than flat. Two differences from flat spacetime matter here. First, de~Sitter spacetime has no null
infinity. Its future boundary $\mathscr I^+$ is spacelike. Second, it has no Killing vector
that is timelike everywhere.  Even the time translation of the static patch of an observer is
spacelike near $\mathscr I^+$. Energy therefore cannot be defined as the charge of an
asymptotic time translation, and the energy carried by gravitational waves is not positive
in general. For these reasons the theory of gravitational radiation changes qualitatively
when $\Lambda>0$.

Ashtekar, Bonga and Kesavan (ABK) \cite{Ashtekar:2015ooa, Ashtekar:2015lxa}
established the framework in the linear theory and derived the quadrupole formula,  building upon the earlier works \cite{deVega:1998ia, Ashtekar:2014zfa, Ashtekar:2015lla}. A decade of
subsequent work has sharpened the picture from several directions \cite{Date:2015kma,  Chrusciel:2016oux, Date:2016uzr, Bishop:2015kay, Hoque:2018byx, Poole:2018koa, Ashtekar:2019khv, Compere:2019bua,   Chrusciel:2020rlz, Compere:2020lrt, Kolanowski:2020wfg,  Chrusciel:2021ttc, Dobkowski-Rylko:2022dva,  Bonga:2023eml, Compere:2023ktn, Chrusciel:2023umn, Harsh:2024kcl, Dobkowski-Rylko:2024jmh, Compere:2024ekl,   Hoque:2025iar, McNees:2025acf, Poole:2025cmv, Harsh:2026vmh}. A  comprehensive treatment of the quadrupolar gravitational radiation was given  by Comp\`ere, Hoque, and Kutluk (CHK)~\cite{Compere:2023ktn}. They derived the metric perturbation at quadrupolar order including both parities, the even (mass quadrupole) and odd (current quadrupole) sectors. In a recent paper, this analysis was extended to octupolar order via a lengthy direct calculation \cite{Harsh:2026vmh}. That extension gave us a number of insights which we  piece together in this paper to derive the gravitational field generated by a localised matter
source in de~Sitter spacetime at arbitrary multipole order. 

The construction rests on two exact structural observations. The Poincar\'e patch
is conformally flat, 
\begin{equation}
d\bar s^2=a^2(\eta)\bigl(-d\eta^2+d\vec x^{\,2}\bigr)
\end{equation}
with
$a=-1/(H\eta)$, so an auxiliary Minkowski metric is naturally present. The source for the de~Sitter problem, which we call the physical stress tensor $T_{\mu\nu}$,  is conserved with respect to the de~Sitter metric. The first observation is that 
an integral of the  spatial part
of the physical stress tensor over the past history of the source  allows us to construct an auxiliary stress tensor
$\Theta_{\mu\nu}$ that is conserved with respect to the auxiliary Minkowski metric,  $\partial^\mu\Theta_{\mu\nu}=0$.\footnote{Indices on flat-space quantities such as $\Theta_{\mu\nu}$ are raised with $\eta^{\mu\nu}$, spatial indices with $\delta_{ij}$.}  The second observation concerns propagation rather than
the source. After a scalar--vector--tensor decomposition and generalised harmonic
gauge fixing, the tensor equation still carries a curvature potential \cite{deVega:1998ia, Ashtekar:2015lxa}. A 
transformation relates its wave operator to the free Minkowski one. The de Sitter ``tail'' is not removed by this step but relocated into the auxiliary source.  The tail is the part of the field that propagates inside the light cone. As in a generic curved
spacetime, the retarded Green function of the tensor equation
is supported inside the past light cone as well as on it, so that Huygens' principle does not hold
and the field depends on the whole past history of the source. In de Sitter spacetime the
violation is very mild. Inside the light cone the Green function is a constant
\cite{deVega:1998ia}, as a result the tail is a single time integral of the stresses of the
source, and this is what allows us to relocate it. Together the two observations turn the de~Sitter problem into the flat-space linearised problem in
harmonic gauge.  Schematically,
\begin{equation*}
T_{\mu\nu}
\;\xrightarrow \;      
\Theta_{\mu\nu}
\;\xrightarrow{\  }\;                         
\text{flat-space field}
\;\xrightarrow{\  }\;      
\text{de Sitter field}.
\end{equation*}
The first and the last arrows involves at most an integral over the past, and the cosmological constant
enters through them. The middle arrow is the retarded solution of the
flat-space linearised Einstein equations in harmonic gauge. Each step is exact. 

This reduction makes the  flat-space  Damour--Iyer multipole results \cite{Damour:1990gj} available.
The developments preceding \cite{Damour:1990gj} are summarised in the classic review of Thorne \cite{Thorne:1980ru}.  The modern multipolar formalism for radiating sources originates in the foundational work of Blanchet and Damour \cite{Blanchet:1985sp} and is comprehensively reviewed by Blanchet \cite{Blanchet:2013haa}. In the earlier de Sitter gravitational waves literature \cite{Date:2015kma, Date:2016uzr, Ashtekar:2015lxa, Compere:2023ktn} the moments are integrals of the
stress tensor over a slice of constant conformal time $\eta$,
\begin{equation}
  Q^{(\rho)}_L=\int\! d^3x'\,a^{\ell+1}T_{00}\,x'_L,\qquad
  P_{i|L}=\int\! d^3x'\,a^{\ell+1}T_{0i}\,x'_L,\qquad
  S_{ij|L}=\int\! d^3x'\,a^{\ell+1}T_{ij}\,x'_L .
  \label{eq:intro-instantaneous-moments}
\end{equation}
Retardation is not carried by these objects. It is restored afterwards, by
Taylor expanding $T_{\mu\nu}(\eta-|\vec x-\vec x\,'|,\vec x\,')$ in the source
size and reorganising the result. The
Damour--Iyer moments are instead radiative from the outset.
They are functions of a single retarded time $\eta_{\rm ret}=\eta-\rho$, $\rho = |\vec x|$, of the field point,
and the source \emph{is smeared along the light cone}. The smearing is implemented via $\delta_\ell(z)$ as follows. Let $\rho'=|\vec x\,'|$,
\begin{equation}
  \eta_z=\eta_{\rm ret}+z\,\rho',\qquad z\in[-1,1],\qquad
  \delta_\ell(z)=\frac{(2\ell+1)!!}{2^{\ell+1}\ell!}\bigl(1-z^2\bigr)^{\ell},
  \qquad \int_{-1}^{1}\delta_\ell(z)\,dz=1 .
  \label{eq:intro-z-average}
\end{equation}
As $z$ sweeps $[-1,1]$, $\eta_z$ sweeps precisely the range of retarded times
over which the past light cone of the field point crosses the source sphere of
radius $\rho'$, and $\delta_\ell$ is the weight that makes the $\ell$-th multipole
exact. The $z$ integral therefore performs the entire finite-retardation average, with no expansion in the source size. The two descriptions are related by
expanding the average,
\begin{equation}
  \int_{-1}^{1}d z\,\delta_\ell(z)\,F(\eta_{\rm ret}+z\rho')
  =F(\eta_{\rm ret})+\frac{\rho'^2}{2(2\ell+3)}\,F^{(2)}(\eta_{\rm ret})+\mathcal O\!\left(\rho'^4\partial_{\eta_{\rm ret}}^4\right),
  \label{eq:intro-z-slow-motion}
\end{equation}
whose leading term collapses the light-cone average onto a constant-time slice.
This reduces to the moments
\eqref{eq:intro-instantaneous-moments} used in the earlier literature.

There is also an organising choice that makes the analysis tractable at arbitrary $\ell$.  The earlier treatments carry the source data in the Cartesian integrals $Q^{(\rho)}_L$, $P_{i|L}$, $S_{ij|L}$, which are reducible under SO(3).  We instead
introduce irreducible symmetric trace-free (STF) radiative moments following
Damour and Iyer \cite{Damour:1990gj} at the outset. Conservation and the residual
harmonic gauge freedom then eliminate the redundant towers algebraically, for all
$\ell$ at once, leaving exactly two families of multipole moments $I_L$ and $J_L$.

With these ideas in place, three things follow immediately. (i) The multipole problem is solved in
one go instead of order by order. (ii)
The relation to the flat-space multipolar formalism becomes direct rather
than an analogy.  (iii) It becomes manifest
that the linearised equations are satisfied separately at each multipolar order,
since each STF tower generates its own exterior solution.  Our main results are the exterior metric at arbitrary 
$\ell \ge 2$, equations~\eqref{eq:h-etaeta}--\eqref{eq:h-ij}, together with equations~\eqref{eq:I-dS}--\eqref{eq:J-dS}, which give $I_L$
 and  $J_L$ as explicit integrals over the physical stress tensor.

The rest of the paper is organised as follows. In Section \ref{sec:eom} we set up the
linearised problem in the Poincar\'e patch and construct the map to flat
spacetime. The Damour--Iyer results apply
verbatim to the mapped problem. Section \ref{sec:poincare} uses this map to construct the
exterior solution at arbitrary multipole order. Section~\ref{sec:comparison} compares with the literature: in the quadrupolar truncation of \cite{Compere:2023ktn},
their metric is our canonical field together with a residual gauge transformation, which we construct explicitly. We conclude in Section \ref{sec:conclusion}. Appendix~\ref{app:stf}
collects our STF conventions and states the Damour--Iyer results in the notation
used here. Appendix \ref{app:retarded} evaluates the metric perturbation in a truncation of the type used in \cite{Compere:2023ktn, Harsh:2026vmh}, made at an arbitrary degree, and compares 
term by term, at degree two with \cite{Compere:2023ktn} and  at degree three with \cite{Harsh:2026vmh}.

\smallskip
\paragraph{Note added:} Shortly after this work was
completed, Comp\`ere and Hoque \cite{Compere:2026xwy} posted a
preprint treating compact sources on spatially flat FLRW backgrounds
with general expansion histories, which obtains the retarded
linearised field at every multipole order together with the
disturbance induced in the cosmic fluid. In pure de Sitter spacetime,
where this fluid response is absent, their results partially overlap
with ours. We comment briefly on the relation between the two
constructions in Section~\ref{sec:conclusion}.

\section{Linearised equations of motion and mapping to flat spacetime}
\label{sec:eom}
In this section we set up the linearised Einstein equations with a source in
the Poincar\'e patch, in generalised harmonic gauge. We then show that they are
equivalent to the flat-space linearised Einstein equations in harmonic gauge,
with a source that is symmetric, spatially compact and conserved.  The equivalence holds for the field equations, the gauge condition, and for the residual gauge freedom. Throughout the paper $\Lambda =3 H^2$, $c=1$, $\Boxf = -\pd_\eta^2+\pd_i^2$. Indices on auxiliary flat-space quantities are raised and lowered with $\eta_{\mu\nu}=\mathrm{diag}(-1,1,1,1)$, and repeated spatial indices are summed with $\delta_{ij}$.

\subsection{Field equations}
The future Poincar\'e patch metric is
\begin{equation}
 {d \bar s}^2=\bar{g}_{\mu \nu}d x^\mu d x^\nu= a^2(\eta)\bigl(-d\eta^2+d\vec x^{\,2}\bigr),\qquad a=-\frac1{H\eta},\qquad \eta<0,
 \label{eq:poincare-metric}
\end{equation}
with $\scri^+$ at $\eta\to0^-$ and the past cosmological horizon at $\eta\to-\infty$. Let $h_{\mu\nu}$ be the metric perturbation and $\widehat h_{\mu\nu}=h_{\mu\nu}-\tfrac12\bar g_{\mu\nu}\bar g^{\alpha\beta}h_{\alpha\beta}$ the corresponding trace-reversed perturbation. We use the rescaled variables
\begin{equation}
 \chi_{\mu\nu}\equiv a^{-2}\widehat h_{\mu\nu},\qquad
 h_{\mu\nu}=a^2\bigl(\chi_{\mu\nu}-\tfrac12\eta_{\mu\nu}\,\eta^{\alpha\beta}\chi_{\alpha\beta}\bigr),
 \label{eq:chi-def}
\end{equation}
and write $\widehat\chi\equiv\chi_{00}+\chi_{kk}$. In components, \eqref{eq:chi-def} gives the line element
\begin{equation}
	ds^2=\bar{g}_{\mu \nu }d x^\mu d x^\nu +a^2\left[\frac{\widehat{\chi}}{2}d \eta^2+2\chi_{0i}d \eta d x^i+\left\{\chi_{ij}+\delta_{ij}\left(\frac{\widehat{\chi}}{2}-\chi_{kk}\right)\right\}d x^i d x^j\right].
    \label{eq:line-element}
\end{equation}
We impose the generalised harmonic gauge condition of \cite{deVega:1998ia,Ashtekar:2015lxa, Compere:2023ktn},
\begin{equation}
 \mathcal C_\mu\equiv\pd^\alpha\chi_{\alpha\mu}+\frac1\eta\bigl(2\chi_{0\mu}+\delta^0_\mu\,\chi^\alpha{}_\alpha\bigr)=0,
 \label{eq:gh-condition}
\end{equation}
that is,
\bea
\label{eq:gh-components1}
 \mathcal C_0 &=&-\pd_\eta\chi_{00}+\pd_i\chi_{0i}+\frac{\widehat\chi}{\eta} =0,\\
 \mathcal C_i&=&-\pd_\eta\chi_{0i}+\pd_j\chi_{ji}+\frac2\eta\chi_{0i}=0.
 \label{eq:gh-components2}
\eea
In this gauge, the linearised Einstein equations with cosmological constant decouple into scalar, vector, and tensor parts \cite{deVega:1998ia}
\begin{align}
	\label{eq:scalarChiEOM}
    \Boxf\left(\frac{\hat{\chi}}{\eta}\right) &=-\frac{16\pi G}{\eta} \widehat{T}, \\
    \label{eq:vectorChiEOM}
	\Boxf\left(\frac{\chi_{0i}}{\eta}\right) &=-\frac{16\pi G }{\eta} T_{0i}, \\
	\left(\Boxf+\frac{2}{\eta^2}\right)\left(\frac{\chi_{ij}}{\eta}\right) &=-\frac{16\pi G }{\eta} T_{ij}. \label{eq:tensorChiEOM}
\end{align}
The scalar and vector equations are flat wave equations. The tensor equation carries the curvature potential $+2/\eta^2$. The corresponding  retarded Green function has support inside the light cone. This is the origin of the de Sitter tail \cite{deVega:1998ia,Ashtekar:2015lxa}.

Let the physical stress tensor sourcing the de Sitter field be $T_{\mu \nu}$. The stress tensor is conserved in de Sitter, $\overline \nabla^\mu T_{\mu\nu}=0$, which in these coordinates reads\footnote{$\overline \nabla_\mu$ is the covariant derivative with respect to the background de Sitter metric $\bar{g}_{\mu \nu}$, and $\overline \nabla^\mu = \bar g^{\mu \nu} \overline \nabla_\nu$.}
\begin{equation}
 \pd_\eta T_{00}=\pd_iT_{0i}+\frac1\eta\widehat T, \label{eq:dS-conservation1}
 \end{equation}
\begin{equation}
 \pd_\eta T_{0i}=\pd_jT_{ij}+\frac2\eta T_{0i},
 \label{eq:dS-conservation2}
\end{equation}
where $\widehat T\equiv T_{00}+T_{kk}$. Throughout the paper the physical stress tensor is taken from the following class.
\begin{itemize}
\item[(S1)] $T_{\mu\nu}(\eta,\vxp)$ is smooth, and for every $\eta$ its support lies in a
fixed comoving ball $\rho'<R_s$. Its physical size is also small compared with the Hubble radius, $a(\eta)\rho'\ll H^{-1}$ on the support of $T_{\mu\nu}$, so that the source is an isolated system rather than a cosmological one. An example is the internal dynamics of a galaxy cluster, where the Hubble expansion plays a role while the cluster, being gravitationally bound, remains small compared with the Hubble radius.
\item[(S2)] The source dies away toward the past cosmological horizon: $T_{\mu\nu}$ and all its
$\eta$-derivatives tend to zero as $\eta\to-\infty$.
\end{itemize}
Two consequences of (S2) are used repeatedly. First, every integral over the past that appears
below converges. Second, the retarded fields constructed below, and the multipole moments, are
built from integrals of the stress tensor over the source and over its past. They therefore
vanish toward the past horizon together with all their $\eta$-derivatives. By the retarded
solution of a field equation we mean, throughout, the solution that vanishes toward the past
horizon.

These conditions are stronger than the assumptions of the quadrupolar and octupolar analyses
\cite{Compere:2023ktn,Harsh:2026vmh}. There the source is spatially compact 
and the only condition placed on its past is that $T_{ij}(\eta,\vec x)$ remain finite as
$\eta\to-\infty$. In particular the source may be stationary in the past. The retarded solutions
of \cite{Compere:2023ktn,Harsh:2026vmh} accordingly contain one constant, $\chi_{ij}(-\infty)$,
which is a limit of moments of the source at the past horizon. For a source in our class the
moments vanish there, so the constant vanishes. Sources that are stationary in the past are therefore not in our class.\footnote{Some of this could be relaxed at the
price of extra terms, such as the static fields of constant charges, or a constant
$\chi_{ij}(-\infty)$, which solves the homogeneous tensor equation and the gauge condition. We
keep the class as it is, to keep the presentation simple.}

This is less restrictive than it appears. The field equations and the gauge
condition are linear, so a source that differs from a stationary one by a stress
tensor in our class is covered by superposition. The stationary part
carries the mass and the angular momentum. These are conserved, and therefore
vanish for a source that dies away toward the past horizon. The conditions
(S1)--(S2) are thus conditions on the time-dependent part of the source alone,
and all the radiation comes from that part. In particular, the sources of~\cite{Ashtekar:2015lxa}, which have bounded
physical size and are stationary outside a finite interval of time, are of
this form. 

With these preliminaries in place, we come to the first of the two observations on which the construction rests. Multiplying the momentum equation in \eqref{eq:dS-conservation2} by $\eta^{-2}$ turns it into a total derivative, $\pd_\eta(T_{0i}/\eta^2)=\pd_j(T_{ij}/\eta^2)$. Define 
\be
 \label{eq:Pi-def}
 \Pi_{ij}(\eta,\vec x)\equiv\int_{-\infty}^{\eta}\frac{\dd\eta'}{\eta'^2}\,T_{ij}(\eta',\vec x),
\ee
so that
\begin{align}
\pd_\eta\Pi_{ij}&=\frac{T_{ij}}{\eta^2}, & & \mathrm{and} & 
\pd_j\Pi_{ij}&=\frac{T_{0i}}{\eta^2}.
\end{align}
The second identity uses (S2) to drop the endpoint term $\eta'^{-2}T_{0i}$ at the past horizon. Note that $\Pi_{ij}$, 
being an integral over the past, records the history of the source. In particular, it does not return to zero after the source switches off. Now set
\bea
  \label{eq:Theta-def1}
&& \Theta_{00}~\equiv~ aT_{00}+\frac1H\Pi_{kk} = a T_{00} + \frac1H \int_{-\infty}^{\eta}\frac{\dd\eta'}{\eta'^2}\,T_{kk}(\eta',\vec x),\\
  \label{eq:Theta-def2}
 && \Theta_{0i}~\equiv~ aT_{0i},\\
&& \Theta_{ij}~\equiv~ aT_{ij}-\frac1H\Pi_{ij}=-\frac1H\,\pd_\eta\bigl(\eta\Pi_{ij}\bigr)= aT_{ij} - \frac1H \int_{-\infty}^{\eta}\frac{\dd\eta'}{\eta'^2}\,T_{ij}(\eta',\vec x),
 \label{eq:Theta-def3}
\eea
Using \eqref{eq:dS-conservation1}--\eqref{eq:dS-conservation2}, one finds
\begin{equation}
\pd^\mu\Theta_{\mu\nu}=0,
 \label{eq:Theta-conserved}
\end{equation}
that is, $\Theta_{\mu \nu}$ is conserved  with respect to the auxiliary Minkowski metric.

Two further properties of $\Theta_{\mu\nu}$ will be used repeatedly. Its trace combination is local, because $\Pi_{kk}$ cancels between $\Theta_{00}$ and $\Theta_{kk}$:
\begin{equation}
 \Theta_{00}+\Theta_{kk}=a\,\widehat T.
 \label{eq:Theta-trace}
\end{equation}
Time derivatives of the spatial part of $\Theta_{\mu\nu}$ are also local in time: 
\begin{equation}
 \pd_\eta\Theta_{ij}=a\,\pd_\eta T_{ij},\qquad
\pd_\eta^2\Theta_{ij}=a\Bigl(\pd_\eta^2-\frac1\eta\pd_\eta\Bigr)T_{ij}.
 \label{eq:Theta-derivatives}
\end{equation}
The overall normalisation is chosen so that $\Theta_{\mu\nu}\to T_{\mu\nu}$ as $H\to0$. To see this introduce the cosmic time coordinate $t$ via $\eta=-H^{-1}e^{-Ht}$. One has $\dd\eta'/\eta'^2=H^2 a(t')\,\dd t'$, so
\[
\Pi_{ij}=H^2\!\int_{-\infty}^{t}\!\dd t'\,a\,T_{ij},\qquad
\Th_{00}=aT_{00}+H\!\int_{-\infty}^{t}\!\dd t'\,a\,T_{kk},\qquad
\Th_{ij}=aT_{ij}-H\!\int_{-\infty}^{t}\!\dd t'\,a\,T_{ij},
\]
which makes $\Th_{\mu\nu}=T_{\mu\nu}+O(H)$ manifest.

We now come to the second observation. With
\be
 \Dm \equiv a^{-1} \partial_\eta a = \pd_\eta-\frac1\eta,\qquad \Dp\equiv a \partial_\eta a^{-1} = \pd_\eta+\frac1\eta,
 \ee
one has
 \be
 \Dp\Dm=\pd_\eta^2,\qquad \Dm\Dp=\pd_\eta^2-\frac2{\eta^2},
\ee
so that the operator in the tensor equation \eqref{eq:tensorChiEOM} satisfies
\be
\left(\Boxf+\frac2{\eta^2}\right)\Dm=\Dm \,\Boxf.
\ee
Define $\Xi_{ij}$ by
\be \label{eq:def_Xi}
\Dm \Xi_{ij} = \frac{\chi_{ij}}{\eta},  \qquad \Xi_{ij}(\eta,\vec x)\equiv\eta\int_{-\infty}^{\eta}\frac{\dd\eta'}{\eta'^2}\,\chi_{ij}(\eta',\vec x), \qquad \chi_{ij} = \eta \Dm \Xi_{ij}.
\ee
The left-hand side of \eqref{eq:tensorChiEOM} then reads
\be
\left(\Boxf+\frac2{\eta^2}\right) \left(\frac{\chi_{ij}}{\eta}\right)= \left(\Boxf+\frac2{\eta^2}\right) \left( \Dm \Xi_{ij}\right) = \Dm \Box \,   \Xi_{ij}. 
\ee
Equating this to  the right-hand side of equation \eqref{eq:tensorChiEOM} gives 
\be
\Dm \Box \,   \Xi_{ij} = -\frac{16\pi G}{\eta}T_{ij},
\ee
which integrates to\footnote{The solution proportional to $\eta$ of the homogeneous equation $D_-f=0$ grows toward the past horizon and is excluded, while $\eta \Pi_{ij}\to0$ by (S2).}
\be
\Box \,   \Xi_{ij} = -16\pi G \,\eta\, \int_{-\infty}^{\eta}\frac{\dd\eta'}{\eta'^2}\,T_{ij}(\eta',\vec x),
\ee
or equivalently 
\be
\label{eq:tensorXiEOM}
\Box \,   \Xi_{ij} = -16 \pi G \, \eta \, \Pi_{ij}(\eta, \vec x), 
\ee
where $\Pi_{ij}$ was defined in \eqref{eq:Pi-def}. 
In this form the curvature potential $+2/\eta^2$ has disappeared from the wave operator. The tail has not been removed, only relocated. It now sits in the time-integrated source $\eta \, \Pi_{ij}(\eta, \vec x)$.

Now comes the key point. Apply to the field the same combination \eqref{eq:Theta-def1}--\eqref{eq:Theta-def3} that builds $\Theta_{\mu\nu}$ out of $T_{\mu\nu}$
\bea
\label{eq:h-of-chi1}
 \mathfrak h_{00}&\equiv&a\,\chi_{00}+\frac1H\int_{-\infty}^{\eta}\frac{\dd\eta'}{\eta'^2}\,\chi_{kk}(\eta',\vec x)
 =a\,\widehat\chi-\mathfrak h_{kk}, 
\\
 \label{eq:h-of-chi2}
 \mathfrak h_{0i}&\equiv&a\,\chi_{0i},\\
 \mathfrak h_{ij}&\equiv&a\,\chi_{ij}-\frac1H\int_{-\infty}^{\eta}\frac{\dd\eta'}{\eta'^2}\,\chi_{ij}(\eta',\vec x),
 \label{eq:h-of-chi3}
\eea
and observe that
\begin{equation}
\Boxf\,\mathfrak h_{\mu\nu}=-16\pi G\,\Theta_{\mu\nu}.
 \label{eq:aux-eq}
\end{equation}
Equation \eqref{eq:aux-eq} is the linearised Einstein equation of flat space in harmonic gauge for the trace-reversed perturbation $\mathfrak h_{\mu\nu}$, with source $\Theta_{\mu\nu}$.
The following relations are useful in verifying \eqref{eq:aux-eq}
\be
\mathfrak{h}_{ij} = - \frac1H \partial_\eta \Xi_{ij}, \qquad [\Box, \partial_\eta]\Xi_{ij} = 0. \label{eq:relations}
\ee

Since $\chi_{ij}$ and $\Xi_{ij}$ vanish toward the past horizon, the field $\chi_{\mu\nu}$ is recovered from $\mathfrak h_{\mu\nu}$ by integrating the first relation in \eqref{eq:relations} and substituting into \eqref{eq:def_Xi}. We get,
\bea
\label{eq:chi-of-h1}
 \chi_{00}&=&a^{-1}\mathfrak h_{00}-H\int_{-\infty}^{\eta}\dd\eta'\,\mathfrak h_{kk}(\eta',\vec x),\\
 \label{eq:chi-of-h2}
 \chi_{0i}&=&a^{-1}\mathfrak h_{0i},\\
 \chi_{ij}&=&a^{-1}\mathfrak h_{ij}+H\int_{-\infty}^{\eta}\dd\eta'\,\mathfrak h_{ij}(\eta',\vec x).
 \label{eq:chi-of-h3}
\eea
From these equations it follows that, 
\begin{equation}
 \widehat\chi=a^{-1}\bigl(\mathfrak h_{00}+\mathfrak h_{kk}\bigr).
 \label{eq:chihat-of-h}
\end{equation}
The de Sitter tail is now explicit. It is the time integral in \eqref{eq:chi-of-h3}. 

Take  $\mathfrak h_{\mu\nu}$ to be the retarded solution
\begin{equation}
 \mathfrak h_{\mu\nu}(\eta,\vec x)=4G\int\dd^3x'\,\frac{\Theta_{\mu\nu}(\eta-|\vec x-\vec x{\,}'|,\vec x{\,}')}{|\vec x-\vec x{\,}'|}.
 \label{eq:h-retarded}
\end{equation}
Our assumptions on the source guarantee that the integral converges. Differentiating \eqref{eq:h-retarded} under the integral, 
and using the conservation law \eqref{eq:Theta-conserved}
one obtains the familiar result
\begin{equation}
 \pd^\mu\mathfrak h_{\mu\nu}=0.
\end{equation}
  Hence $\mathfrak h_{\mu\nu}$ is exactly the object treated by Damour and Iyer \cite{Damour:1990gj}: the retarded field of a symmetric, spatially compact, conserved source.  The field $\chi_{\mu\nu}$ obtained from \eqref{eq:h-retarded} 
    through
\eqref{eq:chi-of-h1}--\eqref{eq:chi-of-h3} is the retarded solution of 
\eqref{eq:scalarChiEOM}--\eqref{eq:tensorChiEOM}.

\subsection{Gauge condition} 
Using these manipulations, we can now solve equations \eqref{eq:scalarChiEOM}--\eqref{eq:tensorChiEOM}. However, 
solving equations \eqref{eq:scalarChiEOM}--\eqref{eq:tensorChiEOM} is not enough, as these equations were derived assuming the gauge condition \eqref{eq:gh-condition}, and a solution that violates it need not solve the Einstein equations. We must therefore check that the harmonic condition on  the auxiliary metric $\mathfrak{h}_{\mu \nu}$ is equivalent to \eqref{eq:gh-condition}.

Write $\mathcal F_\nu\equiv\pd^\mu\mathfrak h_{\mu\nu}=-\pd_\eta\mathfrak h_{0\nu}+\pd_i\mathfrak h_{i\nu}$. For $\nu=0$, we need $\pd_\eta\mathfrak h_{00}$ and $\pd_i\mathfrak h_{i0}$. From \eqref{eq:h-of-chi1}--\eqref{eq:h-of-chi2}, we have, 
\begin{equation}
 \pd_\eta\mathfrak h_{00}=a\,\pd_\eta\chi_{00}-\frac a\eta\chi_{00}+\frac{\chi_{kk}}{H\eta^2}
 =a\Bigl(\pd_\eta\chi_{00}-\frac{\widehat\chi}{\eta}\Bigr),\qquad
 \pd_i\mathfrak h_{i0}=a\,\pd_i\chi_{0i},
\end{equation}
so that, comparing with \eqref{eq:gh-components1},
\begin{equation}
\mathcal F_0=a\Bigl(-\pd_\eta\chi_{00}+\pd_i\chi_{0i}+\frac{\widehat\chi}{\eta}\Bigr)=a\,\mathcal C_0.
 \label{eq:F0}
\end{equation}
Thus $\mathcal C_0=0$ implies $\mathcal F_0 =0$. For $\nu=j$ the computation is slightly longer.  From \eqref{eq:h-of-chi2}--\eqref{eq:h-of-chi3}
\begin{equation}
\label{eq:F_j}
 \mathcal F_j=-a\,\pd_\eta\chi_{0j}+\frac a\eta\chi_{0j}+a\,\pd_i\chi_{ij}
 -\frac1H\int_{-\infty}^{\eta}\frac{\dd\eta'}{\eta'^2}\,\pd_i\chi_{ij}(\eta',\vec x).
\end{equation}
To relate this to \eqref{eq:gh-components2}, rewrite \eqref{eq:gh-components2} as 
\be
\pd_i\chi_{ij}=\mathcal C_j+\pd_\eta\chi_{0j}-\frac2\eta\chi_{0j},
\ee
and insert it into \eqref{eq:F_j}. We get,
\begin{equation}
 \mathcal F_j=a\,\mathcal C_j-\frac1H\int_{-\infty}^{\eta}\frac{\dd\eta'}{\eta'^2}\,\mathcal C_j(\eta',\vec x)
 +\frac1H\lim_{\eta'\to-\infty}\frac{\chi_{0j}(\eta',\vec x)}{\eta'^2}.
 \label{eq:Fj}
\end{equation}
Since $\chi_{0j}\to0$ toward the past horizon by our assumptions, $\mathcal C_j=0$ implies $\mathcal F_j=0$.

The implication also runs the other way. That $\mathcal F_0=0$ implies $\mathcal C_0=0$ is
immediate from \eqref{eq:F0}. For the spatial components, insert
\eqref{eq:chi-of-h2}--\eqref{eq:chi-of-h3} into \eqref{eq:gh-components2}. One finds
\begin{equation}
\mathcal C_j
= a^{-1}\mathcal F_j
+ H\int_{-\infty}^{\eta}\dd\eta'\,\mathcal F_j(\eta',\vec x)
- H\,\mathfrak h_{0j}(-\infty,\vec x).
\label{eq:Cj-of-Fj}
\end{equation}
Hence, if
$\mathcal F_j=0$, then $\mathcal C_j=-H\,\mathfrak h_{0j}(-\infty,\vec x)$ is
independent of $\eta$, and it vanishes precisely when $\mathfrak h_{0j}$ vanishes
toward the past horizon. In other words, on fields that decay toward the past
horizon, the two gauge conditions are equivalent.

\subsection{Residual gauge freedom} 
\label{sec:gauge}
The Damour–Iyer analysis also makes essential use of the residual gauge freedom. For the map to be complete we must therefore show that residual gauge transformations of the auxiliary flat problem correspond to residual gauge transformations of the de Sitter problem. The latter were analysed in \cite{deVega:1998ia, Compere:2023ktn}. We summarise what is needed.  An infinitesimal diffeomorphism $\delta_\xi h_{\mu\nu}=\overline {\nabla}_\mu\xi_\nu+\overline {\nabla}_\nu\xi_\mu$ acts on $\chi_{\mu\nu}$ as the flat-space gauge transformation plus a single curvature term,
\begin{equation}
\delta_\xi\chi_{\mu\nu}=\pd_\mu\xi^\flat_\nu+\pd_\nu\xi^\flat_\mu-\eta_{\mu\nu}\pd_\rho\xi^\rho+\frac2\eta\,\eta_{\mu\nu}\,\xi^0,
 \qquad \xi^\flat_\mu\equiv\eta_{\mu\nu}\xi^\nu.
 \label{eq:dS-gauge-action}
\end{equation}
In components
\bea
&& \delta_\xi\widehat\chi = -4\Dm\xi^0,\\
&& \delta_\xi\chi_{0i}=\pd_\eta\xi^i-\pd_i\xi^0,\\
&& \delta_\xi\chi_{ij}=\pd_i\xi^j+\pd_j\xi^i-\delta_{ij}\bigl(\pd_\eta\xi^0+\pd_k\xi^k\bigr)+\frac2\eta\delta_{ij}\xi^0.
 \label{eq:dS-gauge-components}
\eea
The residual gauge freedom of \eqref{eq:gh-condition} is obtained by evaluating $\mathcal C_\mu$ on \eqref{eq:dS-gauge-action}. The residual gauge vectors satisfy \cite{Compere:2023ktn}
\bea
 \label{eq:residual-dS1}&& \Boxf\left(\frac{\xi^0}{\eta}\right)=0,\\
&& \left(\Boxf+\frac2{\eta^2}\right)\left(\frac{\xi^i}{\eta}\right)=0.
 \label{eq:residual-dS2}
\eea
As for the fields, we require $\xi^\mu$ to vanish, together with its $\eta$-derivatives,
toward the past horizon.

On the flat side the residual gauge freedom consists of vectors $w^\mu$ with $\Boxf w^\mu=0$, which act on the trace-reversed field as
\begin{equation}
 \delta_w\mathfrak{h}_{\mu\nu}=\pd_\mu w_\nu+\pd_\nu w_\mu-\eta_{\mu\nu}\pd_\rho w^\rho,
 \label{eq:flat-gauge-action}
\end{equation}
which implies $\delta_w (\mathfrak h_{00}+\mathfrak h_{kk}) = - 4 \pd_\eta w^0$ and $\delta_w \mathfrak h_{0i}=\pd_\eta w^i-\pd_iw^0$. Requiring that \eqref{eq:chihat-of-h} and \eqref{eq:chi-of-h2} carry $\delta_w\mathfrak h_{\mu\nu}$ into $\delta_\xi\chi_{\mu\nu}$ in the scalar and vector sectors fixes $\xi^\mu$ in terms of $w^\mu$:
\begin{align}
\label{eq:w-to-xi-1}
w^0 &= a \xi^0, & \xi^0 &= a^{-1} w^0 \\
w^i &= \int_{-\infty}^\eta a (\eta') \partial_{\eta'} \xi^i (\eta', \vec x) d \eta', & \xi^i &= a^{-1} w^i + H \int_{-\infty}^\eta w^i (\eta', \vec x) d\eta'.
\label{eq:w-to-xi-2}
\end{align}
For $\xi^\mu$ to vanish toward the past horizon, $a^{-1}w^\mu$ must vanish there and the integrals in \eqref{eq:w-to-xi-2} must converge. This is
the condition we place on $w^\mu$. It remains to check the tensor sector. From \eqref{eq:flat-gauge-action},
\begin{equation}
 \delta_w\mathfrak h_{ij}=\pd_iw^j+\pd_jw^i-\delta_{ij}\bigl(\pd_\eta w^0+\pd_kw^k\bigr).
 \label{eq:dwh-ij-xi}
\end{equation}
Applying the dictionary \eqref{eq:chi-of-h3} to this variation, we have
\begin{equation}
 \delta\chi_{ij}=a^{-1}\,\delta_w\mathfrak h_{ij}+H\int_{-\infty}^{\eta}\dd\eta'\,\delta_w\mathfrak h_{ij}(\eta',\vec x).
\end{equation}
Terms involving the spatial derivative $\pd_iw^j+\pd_jw^i-\delta_{ij}\pd_kw^k$ on the right-hand side of this expression  simplify to, 
\begin{equation}
 a^{-1}\bigl(\pd_iw^j+\pd_jw^i-\delta_{ij}\pd_kw^k\bigr)+H\int_{-\infty}^{\eta}\dd\eta'\,\bigl(\pd_iw^j+\pd_jw^i-\delta_{ij}\pd_kw^k\bigr)
 =\pd_i\xi^j+\pd_j\xi^i-\delta_{ij}\pd_k\xi^k .
\end{equation}
On using $w^0=a\xi^0$ and $w^0\to0$ toward the past horizon, it follows that $\delta\chi_{ij}=\delta_\xi\chi_{ij}$. A similar calculation, using $\Box w^\mu =0$ and the vanishing of $w^\mu$ and $\partial_\eta w^\mu$ towards the past horizon, shows that the vector $\xi^\mu$ so constructed is residual, that is, it satisfies \eqref{eq:residual-dS1}--\eqref{eq:residual-dS2}. Conversely, if $\xi^\mu$ satisfies \eqref{eq:residual-dS1}--\eqref{eq:residual-dS2},
the vector $w^\mu$ satisfies
$\Box w^\mu=0$.

In summary, the complete map is
\begin{equation}
 T_{\mu\nu}\ \xrightarrow{\ \eqref{eq:Theta-def1}-\eqref{eq:Theta-def3}\ }\ \Theta_{\mu\nu}
 \ \xrightarrow{\ \text{flat retarded via }\ \eqref{eq:h-retarded}\ }\ \mathfrak h_{\mu\nu}
 \ \xrightarrow{\ \eqref{eq:chi-of-h1}-\eqref{eq:chi-of-h3}\ }\ \chi_{\mu\nu},
 \label{eq:logic-chain}
\end{equation}
and $\chi_{\mu\nu}$ solves \eqref{eq:scalarChiEOM}--\eqref{eq:tensorChiEOM} in the gauge \eqref{eq:gh-condition}. The residual harmonic gauge freedom of the auxiliary problem is realised by residual de Sitter diffeomorphisms and vice versa.

\section{Poincar\'e patch solutions}
\label{sec:poincare}
In Section~\ref{sec:eom}  we  turned the sourced de Sitter spacetime problem into a sourced flat spacetime problem. What remains is  (i) to solve the flat problem in the exterior of the source, (ii) to organise the solution into irreducible multipole moments, and (iii) to map it back. The flat spacetime multipole results  of Damour and Iyer do the first two steps for all $\ell$ at once. This section carries out the three steps. Section~\ref{sec:direct} writes the retarded solutions of the three sectors directly, and introduces the light-cone moments and the reverse Bessel polynomials. Section~\ref{sec:moments} applies the Damour--Iyer reduction to $\Th_{\mu\nu}$ and obtains the two families of moments $I_L$, $J_L$ as light-cone integrals of the physical stress tensor. Section~\ref{sec:metric} reconstructs the physical metric and states the all-$\ell$ result. Section~\ref{sec:flat} passes to cosmic time and takes the flat limit $H\to0$,
in which the moments and the canonical metric reduce to their flat-space
Damour--Iyer forms.

\subsection{Retarded solutions}
\label{sec:direct}

The scalar and vector equations \eqref{eq:scalarChiEOM}--\eqref{eq:vectorChiEOM} are flat wave equations for $\widehat{\chi}/\eta$ and $\chi_{0i}/\eta$, and the tensor equation \eqref{eq:tensorXiEOM} is a flat wave equation for $\Xi_{ij}$. Their retarded solutions, with the boundary condition that all fields vanish toward the past horizon, are therefore
\begin{align}
\chihat(\eta,\vx)&=4G\int\dd^3x'\;\frac{\eta}{\eta-|\vx-\vxp|}\;\frac{\widehat{T}(\eta-|\vx-\vxp|,\vxp)}{|\vx-\vxp|},
\label{eq:direct-scalar}\\
\chi_{0i}(\eta,\vx)&=4G\int\dd^3x'\;\frac{\eta}{\eta-|\vx-\vxp|}\;\frac{T_{0i}(\eta-|\vx-\vxp|,\vxp)}{|\vx-\vxp|},
\label{eq:direct-vector}\\
\Xi_{ij}(\eta,\vx)&=4G\int\dd^3x'\;\frac{\big[\eta'\,\Pi_{ij}(\eta',\vxp)\big]_{\eta'=\eta-|\vx-\vxp|}}{|\vx-\vxp|}.
\label{eq:direct-Xi}
\end{align}
Since $\chi_{ij}=\eta\,\Dm\Xi_{ij}$,  carrying out the derivative in  \eqref{eq:direct-Xi}, with $\pd_\eta\Pi_{ij}=T_{ij}/\eta^2$, we can write 
\begin{equation}
\chi_{ij}(\eta,\vx)=4G\int\dd^3x'\left\{
\frac{T_{ij}(\eta-|\vx-\vxp|,\vxp)}{|\vx-\vxp|}
+\int_{-\infty}^{\eta-|\vx-\vxp|}\dd\eta'\;\frac{\pd_{\eta'}T_{ij}(\eta',\vxp)}{\eta'}\right\},
\label{eq:direct-tensor}
\end{equation}
for a source decaying toward the past horizon. This is equation (2.80) of CHK \cite{Compere:2023ktn}. The first term of \eqref{eq:direct-tensor} is the flat retarded integral.  The second is supported inside the past light cone of the field point and is the de Sitter tail.

Every retarded integral above is the retarded solution of a flat wave equation, so the scalar result \eqref{eq:scalar-theorem} applies to each of these integrals. In the exterior, the solution of $\Box V=-4\pi S$ is 
\be \label{eq:scalar-expansion}
V(\eta,\vec x)=\sum_{\ell\ge0}\frac{(-1)^\ell}{\ell!}\,\pd_L\!\left[\frac{F_L(\etaret)}{\rho}\right],
\ee
with 
\be
F_L(\etaret)=\int\dd^3x'\,\hxp_L\int_{-1}^1 \dd z\,\delta_\ell(z)S(\etaz,\vxp).
\ee
As shown in the appendix, this expression can also be written as
\be
\label{eq:F_L-final-main-text}
F_L(\etaret)=\int\dd^3x'\;\widehat{x}{\,}'_L\;{}_0F_1\!\Big(\ell+\tfrac32;\tfrac14\rho'^{\,2}\pd_{\etaret}^2\Big)S(\etaret, \vec x{\,}').
\ee

The exterior multipole waves are most compactly written with the reverse Bessel polynomial \cite[\href{http://dlmf.nist.gov/18.34.E2}{18.34.2}]{NIST:DLMF} (see also \cite{Loganayagam:2025jmw})
\begin{align}
\label{eq:theta}
\theta_\ell(z)&=\sum_{k=0}^{\ell}\frac{(\ell+k)!}{2^k\,k!\,(\ell-k)!}\,z^{\ell-k}
=z^\ell+\frac{\ell(\ell+1)}{2} z^{\ell-1}+\cdots+(2\ell-1)!!,\\
\theta_{\ell+1}&=(2\ell+1)\theta_\ell+z^2\theta_{\ell-1},
\qquad
h^{(1)}_\ell(x)=-\mathrm i\,\frac{e^{\mathrm ix}}{x^{\ell+1}}\,\theta_\ell(-\mathrm ix),
\label{eq:theta2}
\end{align}
so that $\theta_\ell$ is the polynomial that turns the outgoing spherical Hankel function into $e^{\mathrm ix}/x^{\ell+1}$. Its role here is the identity
\begin{equation}
\frac{(-1)^\ell}{\ell!}\,\pd_L\!\left[\frac{F_L(\etaret)}{\rho}\right]
=\frac{1}{\ell!}\,\frac{\widehat{n}_L}{\rho^{\ell+1}}\;\theta_\ell(\rho\,\pd_{\etaret})\,F_L(\etaret),
\label{eq:theta-wave}
\end{equation}
for any STF $F_L$,
with 
\be
\theta_\ell(\rho\pd_{\etaret})F_L(\etaret)=\sum_{k=0}^{\ell}\frac{(\ell+k)!}{2^k \, k!\,(\ell-k)!}\rho^{\ell-k}F_L^{(\ell-k)}(\etaret).
\ee 
In  expression \eqref{eq:theta-wave}, the $k=0$ term is the  radiative ($1/\rho$) term with $\ell$ time derivatives, the $k=\ell$ term is the  ($1/\rho^{\ell+1}$) term  with no time derivatives. The scalar expansion \eqref{eq:scalar-expansion} becomes 
\be
\label{eq:scalar-theta_2}
 V(\eta,\vec x)=\sum_{\ell\ge0}\frac{1}{\ell!}\,\frac{\widehat{n}_L}{\rho^{\ell+1}}\;\theta_\ell(\rho\,\pd_{\etaret})\,F_L(\etaret).
 \ee

Applied to \eqref{eq:direct-scalar}--\eqref{eq:direct-Xi}, we get all expansions in one notation, 
\begin{align}
\label{eq:direct-STF}
&\chihat=4G\,\eta\sum_{\ell\ge0}\frac{1}{\ell!}\frac{\hn_L}{\rho^{\ell+1}}\,\theta_\ell(\rho\pd_{\etaret})\,\fQ_L(\etaret),
\\
&\chi_{0i}=4G\,\eta\sum_{\ell\ge0}\frac{1}{\ell!}\frac{\hn_L}{\rho^{\ell+1}}\,\theta_\ell(\rho\pd_{\etaret})\,\fP_{i|L}(\etaret),
\label{eq:direct-STF-sv}\\
&\Xi_{ij}=4G\sum_{\ell\ge0}\frac{1}{\ell!}\frac{\hn_L}{\rho^{\ell+1}}\,\theta_\ell(\rho\pd_{\etaret})\,\fS_{ij|L}(\etaret),
\label{eq:direct-STF-t}
\end{align}
with the light-cone moments
\begin{align}
\label{eq:Q-moments}
&\fQ_L=\int\dd^3x'\,\hxp_L\;\cS_\ell\!\left[\frac{\That}{\eta}\right], \\
&\fP_{i|L}=\int\dd^3x'\,\hxp_L\;\cS_\ell\!\left[\frac{T_{0i}}{\eta}\right],\\
&\fS_{ij|L}=\int\dd^3x'\,\hxp_L\;\cS_\ell\left[\eta\,\Pi_{ij}\right],
\label{eq:QPS-moments}
\end{align}
all evaluated at $\etaret$. Here we have also introduced a convenient name for the light-cone average that defines the moments
\begin{equation}
\cS_\ell[F](\etaret,\vxp) \equiv
\int_{-1}^{1}\dd z\,\delta_\ell(z)F(\etaret+z\rho',\vec x') =
{}_0F_1\!\left(\ell+\tfrac32;\tfrac14\rhop^{\,2}\pd_{\etaret}^2\right)F(\etaret,\vxp).
\label{eq:S-op}
\end{equation}
The vertical bar separates the tensor index from the STF multi-index. The moments $\fP_{i|L}$ and $\fS_{ij|L}$ are STF in $L$ only.

The three families $\fQ_L$, $\fP_{i|L}$, $\fS_{ij|L}$ are not the physical moments. $\fP_{i|L}$ and $\fS_{ij|L}$ carry free indices and are reducible under rotations, and the three families are related by the conservation law, so the exterior field depends on far fewer functions than equations  \eqref{eq:direct-STF}--\eqref{eq:direct-STF-t} suggest. Isolating the physical content by hand is what the quadrupolar and octupolar analyses had to do order by order \cite{Compere:2023ktn, Harsh:2026vmh}.

\subsection{Exact source moments}
\label{sec:moments}

The point of mapping to the auxiliary problem is that the Damour--Iyer reduction can be imported for all $\ell$ at once. It is $\fh_{\mu\nu}$ and $\Th_{\mu\nu}$, not $\chi_{\mu\nu}$ and $T_{\mu\nu}$, that the Damour--Iyer reduction applies to. In those variables the  raw expansion is \eqref{eq:raw-expansion}--\eqref{eq:raw-expansion2}:
\bea
&&
\mathfrak h^{\mu\nu}(\eta,\vec x)=4G\sum_{\ell\ge0}\frac{(-1)^\ell}{\ell!}\,\pd_L\!\left[\frac{F^{\mu\nu}_L(\etaret)}{\rho}\right]  =4G\sum_{\ell\ge0}\frac{1}{\ell!}\frac{\hn_L}{\rho^{\ell+1}}\,\theta_\ell(\rho\pd_{\etaret})\,F^{\mu\nu}_L(\etaret),
\label{eq:raw-expansion-main-text} \\
&& F^{\mu\nu}_L(\etaret)=\int\dd^3x'\;\widehat{x}{\,}'_L \, \cS_\ell  \left[\Theta^{\mu\nu}\right](\etaret, \vxp). 
\label{eq:raw-expansion2-main-text}
\eea

The tensors $F^{0i}_L$ and $F^{ij}_L$ introduced in \eqref{eq:raw-expansion-main-text} are STF only in $L$ and are therefore reducible. A key result of Damour and Iyer \cite{Damour:1990gj} is that conservation and the residual harmonic gauge freedom reduce the ten independent STF families to two. For $\ell\ge0$ and $\ell\ge1$ respectively they define:
\begin{align}
I_L(\etaret)&= \int\dd^3x'\int_{-1}^{1}\dd z\;\Big\{
\delta_\ell\,\widehat{x}{\,}'_L\,\widetilde\sigma
-A_\ell\,\delta_{\ell+1}\,\widehat{x}{\,}'_{aL}\,\pd_{\etaret}\widetilde\sigma_a
+B_\ell\,\delta_{\ell+2}\,\widehat{x}{\,}'_{abL}\,\pd^2_{\etaret}\widetilde\sigma_{ab}\Big\},
\label{eq:I-L-main-tex}\\
J_L(\etaret)&=\STF_L\int\dd^3x'\int_{-1}^{1}\dd z\;\Big\{
\delta_\ell\,\epsilon_{i_\ell ab}\,x'_a\,\widehat{x}{\,}'_{L-1}\,\widetilde\sigma_b
-C_\ell\,\delta_{\ell+1}\,\epsilon_{i_\ell ab}\,\widehat{x}{\,}'_{acL-1}\,\pd_{\etaret}\widetilde\sigma_{bc}\Big\},
\label{eq:J-L-main-tex}
\end{align}
with
\begin{align}
\sigma &= \Theta_{00}+ \Theta_{kk}, &
\sigma_a &= -\Theta_{0a}, &
\sigma_{ab} &= \Theta_{ab},
\end{align}
and tilde denotes evaluation at the light-cone time of the source point
\begin{align}
\widetilde\sigma &\equiv\sigma(\eta_z,\vec x{\,}'),&
\widetilde\sigma_a&\equiv\sigma_a(\eta_z,\vec x{\,}'),&
\widetilde\sigma_{ab}&\equiv\sigma_{ab}(\eta_z,\vec x{\,}'),&
\eta_z=\etaret+z\rho',
\label{eq:tilde-main-tex}
\end{align}
together with
\begin{equation}
A_\ell=\frac{4(2\ell+1)}{(\ell+1)(2\ell+3)},\qquad
B_\ell=\frac{2(2\ell+1)}{(\ell+1)(\ell+2)(2\ell+5)},\qquad
C_\ell=\frac{2\ell+1}{(\ell+2)(2\ell+3)}.
\label{eq:ABC-main-tex}
\end{equation}

The canonical exterior field is determined by $I_L$ and $J_L$ alone:
\begin{equation}
\mathfrak h_{\mu\nu}=\mathfrak h^{\rm can}_{\mu\nu}+\pd_\mu w_\nu+\pd_\nu w_\mu-\eta_{\mu\nu}\,\pd_\alpha w^\alpha,
\qquad \Box w_\mu=0\quad(\rho>R_s),
\label{eq:theorem-main-text}
\end{equation}
where the canonical field is
\begin{align}
\mathfrak h^{\rm can}_{00}&=4G\sum_{\ell\ge0}\frac{(-1)^\ell}{\ell!}\;\pd_L\!\left[\frac{I_L(\etaret)}{\rho}\right],
\label{eq:can-00-main-tex}\\
\mathfrak h^{\rm can}_{0i}&=4G\sum_{\ell\ge1}\frac{(-1)^\ell}{\ell!}\left\{
\pd_{L-1}\!\left[\frac{\dot I_{iL-1}(\etaret)}{\rho}\right]
+\frac{\ell}{\ell+1}\,\epsilon_{iab}\,\pd_{aL-1}\!\left[\frac{J_{bL-1}(\etaret)}{\rho}\right]\right\},
\label{eq:can-0i-main-tex}\\
\mathfrak h^{\rm can}_{ij}&=4G\sum_{\ell\ge2}\frac{(-1)^\ell}{\ell!}\left\{
\pd_{L-2}\!\left[\frac{\ddot I_{ijL-2}(\etaret)}{\rho}\right]
+\frac{2\ell}{\ell+1}\,\pd_{aL-2}\!\left[\frac{\epsilon_{ab(i}\dot J_{j)bL-2}(\etaret)}{\rho}\right]\right\}.
\label{eq:can-ij-main-tex}
\end{align}
A dot  on a function of $\etaret$ denotes $\dd/\dd\etaret$. Four further STF families $W_L,X_L,Y_L,Z_L$ that enter $w_\mu$  are not needed for the canonical field; they enter only the residual gauge vector, which we use in Section~\ref{sec:comparison}.

The moments \eqref{eq:I-L-main-tex}--\eqref{eq:J-L-main-tex} are integrals of the auxiliary source. Inserting \eqref{eq:Theta-def1}--\eqref{eq:Theta-def3} we have for the quantities that enter the definition of these moments in terms of the physical stress tensor
\begin{align}
\label{eq:Sigma-dS}
&\sigma=\Th_{00}+\Th_{kk}=a\,\That,\\
&\sigma_i=-\Th_{0i}=-a\,T_{0i},\\
&\sigma_{ij}=\Th_{ij}=a\,T_{ij}-\frac1H\,\Pi_{ij},\\
&\pd_\eta\sigma_{ij}=a\,\pd_\eta T_{ij}=\pd_\eta\big(aT_{ij}\big)+\frac{a}{\eta}\,T_{ij},\\
&\pd_\eta^2\sigma_{ij}=a\Big(\pd_\eta^2-\frac1\eta\pd_\eta\Big)T_{ij}=\pd_\eta\big(a\,\pd_\eta T_{ij}\big),\\
&\pd_\eta\sigma_i=-\pd_\eta\big(aT_{0i}\big)=-a\Big(\pd_\eta-\frac1\eta\Big)T_{0i}.
\label{eq:derivative-dictionary}
\end{align}
Substituting \eqref{eq:Sigma-dS}--\eqref{eq:derivative-dictionary} in \eqref{eq:I-L-main-tex}--\eqref{eq:J-L-main-tex} gives the moments as light-cone integrals of the physical stress tensor
\begin{align}
I_L(\etaret)&=\int\dd^3x'\int_{-1}^{1}\dd z\;a(\etaz)\Big\{
\delta_\ell(z)\,\hxp_L\,\That
+A_\ell\,\delta_{\ell+1}(z)\,\hxp_{aL}\Big(\pd_\eta-\frac1\eta\Big)T_{0a}\nonumber\\
&\hspace{9.5em}+B_\ell\,\delta_{\ell+2}(z)\,\hxp_{abL}\Big(\pd_\eta^2-\frac1\eta\pd_\eta\Big)T_{ab}
\Big\}_{(\etaz,\vxp)},
\label{eq:I-dS}\\
J_L(\etaret)&=-\STF_L\int\dd^3x'\int_{-1}^{1}\dd z\;a(\etaz)\Big\{
\delta_\ell(z)\,\epsilon_{i_\ell ab}\,x'_a\hxp_{L-1}\,T_{0b}
\nonumber\\
&\hspace{9.5em}+C_\ell\,\delta_{\ell+1}(z)\,\epsilon_{i_\ell ab}\,\hxp_{acL-1}\,\pd_\eta T_{bc}
\Big\}_{(\etaz,\vxp)},
\label{eq:J-dS}
\end{align}
where all stress-tensor components, their $\eta$-derivatives, and $1/\eta$  are evaluated at the light-cone time $\etaz=\etaret+z\rhop$ of the source point and $A_\ell, B_\ell, C_\ell$ are given in \eqref{eq:ABC-main-tex}. These expressions are the first main result of this paper.

Several comments are in order: 
\begin{enumerate}
    \item The moments are integrals of $T_{\mu\nu}$ and its first two time derivatives over the light-cone slice of the source, exactly as in flat space.
    \item Relative to the flat moments the de Sitter moments differ by the conformal factor $a(\etaz)$ under the integral and by the replacement of $\pd_\eta$ acting on $T_{0a}$ and $T_{ab}$ by the combinations in \eqref{eq:Sigma-dS}--\eqref{eq:derivative-dictionary}.  
    \item Only the trace-free part of $T_{ab}$ enters the $B_\ell$ and $C_\ell$ terms, because $\hxp_{abL}$ is trace-free and $\epsilon_{i_\ell ab}\hxp_{abL-1}=0$. The trace $T_{kk}$ enters only through $\That$.
\end{enumerate}

\subsection{The exterior metric at arbitrary \texorpdfstring{$\ell$}{l}}
\label{sec:metric}

We now map the canonical field back to de Sitter. The inverse map
\eqref{eq:chi-of-h1}--\eqref{eq:chihat-of-h} is algebraic in the scalar and
vector sectors, but its tensor part \eqref{eq:chi-of-h3} contains a time
integral, which could reintroduce the past history of the source into the
physical metric. It does not, because of two properties of the canonical
tensor field \eqref{eq:can-ij-main-tex}. It is trace-free
and it is an exact $\eta$-derivative,
\bea
\mathfrak h^{\rm can}_{ij}&=&\pd_\eta Q^{\rm can}_{ij},\\
Q^{\rm can}_{ij}&\equiv& 4G\sum_{\ell\ge2}\frac{(-1)^\ell}{\ell!}\left\{
\pd_{L-2}\!\left[\frac{\dot I_{ijL-2}(\etaret)}{\rho}\right]
+\frac{2\ell}{\ell+1}\,\pd_{aL-2}\!\left[\frac{\epsilon_{ab(i}J_{j)bL-2}(\etaret)}{\rho}\right]\right\},
\label{eq:Q}
\eea
which is the canonical tensor structure with one time derivative fewer on each
moment, and is trace-free for the same reason, $Q^{\rm can}_{kk}=0$. In the scalar and vector sectors, $\fh^{\rm can}_{kk}=0$ gives
\be
\chihat^{\rm can}=a^{-1}\fh^{\rm can}_{00}
\ee
and 
\be
\chi^{\rm can}_{0i}=a^{-1}\fh^{\rm can}_{0i}.
\ee
In the tensor sector the
exact-derivative form pays off:
\begin{equation}
\chi_{ij}^{\rm can}=a^{-1}\pd_\eta Q_{ij}^{\rm can}+H\Big[Q_{ij}^{\rm can}(\eta,\vec x)-Q_{ij}^{\rm can}(-\infty,\vec x)\Big]
=a^{-1}\pd_\eta Q_{ij}^{\rm can}+H\,Q_{ij}^{\rm can},
\label{eq:chi-ij}
\end{equation}
where the endpoint vanishes because the moments are light-cone integrals of a
source that dies away toward the past horizon, so that $I_L(\etaret)$ and
$J_L(\etaret)$ and all their $\eta$-derivatives vanish as $\etaret\to-\infty$. Since $Q^{\rm can}_{kk}=0$, the
canonical $\chi^{\rm can}_{ij}$ has vanishing spatial trace, $\chi^{\rm can}_{kk}=0$.

Inserting $\chihat^{\rm can}$, $\chi_{0i}^{\rm can}$, $\chi_{ij}^{\rm can}$ into the line element \eqref{eq:line-element} gives the physical metric perturbation in the exterior of the source
\begin{align}
h^{\rm can}_{\eta\eta}&=\frac a2\,\mathfrak h ^{\rm can}_{00}
=2Ga\sum_{\ell\ge0}\frac{(-1)^\ell}{\ell!}\;\pd_L\!\left[\frac{I_L}{\rho}\right],
\label{eq:h-etaeta}\\
h^{\rm can}_{\eta i}&=a\,\mathfrak h^{\rm can}_{0i}
=4Ga\sum_{\ell\ge1}\frac{(-1)^\ell}{\ell!}\left\{
\pd_{L-1}\!\left[\frac{\dot I_{iL-1}}{\rho}\right]
+\frac{\ell}{\ell+1}\,\epsilon_{iab}\,\pd_{aL-1}\!\left[\frac{J_{bL-1}}{\rho}\right]\right\},
\label{eq:h-etai}\\
h^{\rm can}_{ij}&=\frac a2\,\delta_{ij}\,\mathfrak h^{\rm can}_{00}+a\,\mathfrak h^{\rm can}_{ij}+H a^2\,Q_{ij}^{\rm can}\nonumber\\
&=2Ga\,\delta_{ij}\sum_{\ell\ge0}\frac{(-1)^\ell}{\ell!}\pd_L\!\left[\frac{I_L}{\rho}\right] \nonumber\\
&\quad+4Ga\sum_{\ell\ge2}\frac{(-1)^\ell}{\ell!}\left\{
\pd_{L-2}\!\left[\frac{\ddot I_{ijL-2}}{\rho}\right]
+\frac{2\ell}{\ell+1}\,\pd_{aL-2}\!\left[\frac{\epsilon_{ab(i}\dot J_{j)bL-2}}{\rho}\right]\right\}\nonumber\\
&\quad+4GHa^2\sum_{\ell\ge2}\frac{(-1)^\ell}{\ell!}\left\{
\pd_{L-2}\!\left[\frac{\dot I_{ijL-2}}{\rho}\right]
+\frac{2\ell}{\ell+1}\,\pd_{aL-2}\!\left[\frac{\epsilon_{ab(i}J_{j)bL-2}}{\rho}\right]\right\}.
\label{eq:h-ij}
\end{align}
These expressions are the second main result of this paper.

The moments $I_L$ and $J_L$ determine the canonical field \eqref{eq:h-etaeta}--\eqref{eq:h-ij}, and the residual gauge vector is
built from four further STF families $W_L$, $X_L$, $Y_L$, $Z_L$, given in equations \eqref{eq:W-aux}--\eqref{eq:Z-aux}. These determine the gauge transformation needed to reconstruct the raw field \eqref{eq:raw-expansion-main-text} from the canonical one and vice versa. For completeness we record them in terms
of the physical stress tensor as well, by the same substitution \eqref{eq:Sigma-dS}--\eqref{eq:derivative-dictionary}:
\begin{align}
W_L&=-\int d^3x'\int_{-1}^{1}dz\;a(\eta_z)\left\{\frac14 A_\ell\,\delta_{\ell+1}(z)\,\widehat{x}{\,}'_{aL}\,T_{0a}
+\frac14 B_\ell\,\delta_{\ell+2}(z)\,\widehat{x}{\,}'_{abL}\,\partial_\eta T_{ab}\right\}_{(\eta_z,\vec{x}\,')},
\label{eq:W-dS}\\[6pt]
X_L&=\frac14 B_\ell\int d^3x'\int_{-1}^{1}dz\;\delta_{\ell+2}(z)\,\widehat{x}{\,}'_{abL}\left\{a\,T_{ab}-\frac{1}{H}\,\Pi_{ab}\right\}_{(\eta_z,\vec{x}\,')},
\label{eq:X-dS}\\[6pt]
Y_L&=-\int d^3x'\int_{-1}^{1}dz\;\Bigg\{\delta_\ell(z)\,\widehat{x}{\,}'_L\Big(a\,T_{aa}-\frac{1}{H}\,\Pi_{aa}\Big)
\nonumber\\
&\hspace{4cm}
+a \Bigg[\frac34 A_\ell\,\delta_{\ell+1}(z)\,\widehat{x}{\,}'_{aL}\Big(\partial_\eta-\frac{1}{\eta}\Big)T_{0a}
\nonumber\\
&\hspace{4.4cm}
+B_\ell\,\delta_{\ell+2}(z)\,\widehat{x}{\,}'_{abL}\Big(\partial_\eta^2-\frac{1}{\eta}\,\partial_\eta\Big)T_{ab}\Bigg]\Bigg\}_{(\eta_z,\vec{x}\,')},
\label{eq:Y-dS}\\[6pt]
Z_L&=C_\ell\,\mathrm{STF}_L\int d^3x'\int_{-1}^{1}dz\;\delta_{\ell+1}(z)\,\epsilon_{i_\ell ab}\,\widehat{x}{\,}'_{acL-1}\left\{a\,T_{bc}-\frac{1}{H}\,\Pi_{bc}\right\}_{(\eta_z,\vec{x}\,')}.
\label{eq:Z-dS}
\end{align}
As in \eqref{eq:I-dS}--\eqref{eq:J-dS}, all stress-tensor components, their $\eta$-derivatives, $\Pi_{ab}$ and $1/\eta$ are evaluated at the
light-cone time $(\eta_z,\vec{x}\,')$ of the source point.  $W_L$ and $X_L$ are defined for $\ell\ge0$, $Y_L$ and $Z_L$ for
$\ell\ge1$. $I_L$, $J_L$ and $W_L$ contain $T_{\mu\nu}$ and its first two
$\eta$-derivatives only. $X_L$, $Y_L$ and $Z_L$ contain $\Pi_{ab}$ undifferentiated, and therefore an integral over the entire past history of the source. For a source in our class the residual gauge vector built from these families satisfies the
conditions on gauge vectors stated in Section~\ref{sec:gauge}.

\subsection{Flat limit}
\label{sec:flat}
We close the section with a discussion of the flat-space limit $H\to0$. In the $(\eta,\vx)$
coordinates this limit is not available. It becomes available in cosmic time coordinates $(t,\vx)$,
defined by
\be
\eta=-\frac1H\,e^{-Ht},\qquad a=e^{Ht}.
\label{eq:cosmic-time}
\ee
Throughout this subsection $T_{00}$, $T_{0i}$ and $T_{ij}$ continue to denote
the components of the stress tensor in the $(\eta,\vec x)$ chart, now regarded as
functions of $t$. They are related to the cosmic-time components by
$T_{\eta\eta}=a^2T_{tt}$, $T_{\eta i}=a\,T_{ti}$, with $T_{ij}$ unchanged. Since
$a\to1$ at fixed $t$ as $H\to0$, the distinction disappears in the limit.  The two derivative combinations that appear in the moments \eqref{eq:I-dS}--\eqref{eq:J-dS} are
\be
\pd_\eta-\frac1\eta=a\big(\pd_t+H\big),\qquad
\pd_\eta^2-\frac1\eta\,\pd_\eta=a^2\big(\pd_t^2+2H\,\pd_t\big),
\label{eq:derivatives-t}
\ee
which have a smooth $H\to0$ limit. In the same variables the conservation equations \eqref{eq:dS-conservation1}--\eqref{eq:dS-conservation2} read
\begin{equation}
\partial_t T_{00}=e^{-Ht}\,\partial_iT_{0i}-H\,\widehat{T},\qquad
\partial_t T_{0i}=e^{-Ht}\,\partial_jT_{ij}-2H\,T_{0i},
\label{eq:conservation-t}
\end{equation}
which in the $H\to0$ limit become $\partial_tT_{00}=\partial_iT_{0i}$, $\partial_tT_{0i}=\partial_jT_{ij}$.

In order to take the $H\to0$ limit of  the moments
\eqref{eq:I-dS}--\eqref{eq:J-dS} we must now convert these expressions to 
the cosmic time coordinates $(t,\vx)$.  With $\eta_{\mathrm{ret}}=-1/(Ha_{\mathrm{ret}})$ and $a_{\mathrm{ret}}=e^{Ht_{\mathrm{ret}}}$, the light-cone time of the source point, $\eta_z=\eta_{\mathrm{ret}}+z\rho'=-H^{-1}e^{-Ht_z}$, corresponds to
\begin{equation}
a(\eta_z)=e^{Ht_z}=\frac{a_{\mathrm{ret}}}{1-Ha_{\mathrm{ret}}\,z\rho'},\qquad
t_z=t_{\mathrm{ret}}-\frac{1}{H}\ln\big(1-Ha_{\mathrm{ret}}\,z\rho'\big).
\label{eq:tz}
\end{equation}
Note that  $t_z$ is regular at $H=0$: $t_z=t_{\mathrm{ret}}+a_{\mathrm{ret}}z\rho'+\tfrac12 Ha_{\mathrm{ret}}^2z^2\rho'^2+O(H^2)$.

Inserting (\ref{eq:derivatives-t})--(\ref{eq:tz}) into \eqref{eq:I-dS}--\eqref{eq:J-dS} gives
\begin{align}
I_L(\eta_{\mathrm{ret}})&=\int d^3x'\int_{-1}^{1}dz\;e^{Ht_z}\Big\{\delta_\ell(z)\,\widehat{x}{\,}'_L\,\widehat{T}
+A_\ell\,\delta_{\ell+1}(z)\,\widehat{x}{\,}'_{aL}\,e^{Ht}\big(\partial_t+H\big)T_{0a}
\nonumber\\
&\hspace{4.2cm}
+B_\ell\,\delta_{\ell+2}(z)\,\widehat{x}{\,}'_{abL}\,e^{2Ht}\big(\partial_t^2+2H\,\partial_t\big)T_{ab}\Big\}_{(t_z,\vec{x}\,')},
\label{eq:I-dS-tret}\\[6pt]
J_L(\eta_{\mathrm{ret}})&=-\mathrm{STF}_L\int d^3x'\int_{-1}^{1}dz\;e^{Ht_z}\Big\{\delta_\ell(z)\,\epsilon_{i_\ell ab}\,x'_a\,\widehat{x}{\,}'_{L-1}\,T_{0b}
\nonumber\\
&\hspace{4.2cm}
+C_\ell\,\delta_{\ell+1}(z)\,\epsilon_{i_\ell ab}\,\widehat{x}{\,}'_{acL-1}\,e^{Ht}\,\partial_t T_{bc}\Big\}_{(t_z,\vec{x}\,')}.
\label{eq:J-dS-tret}
\end{align}
The right-hand sides depend on $\eta_{\mathrm{ret}}$ only through $t_{\mathrm{ret}}$. No inverse power of $H$ appears except inside $t_z$,
which is regular at $H=0$.

In the limit $H\to0$ at fixed $t_{\mathrm{ret}}$, $z$ and $\vec{x}\,'$, and for stress tensors $T_{\mu\nu}(t,\vec{x}\,')$ with a smooth limit, $a_{\mathrm{ret}}\to1$, $e^{Ht_z}\to1$ and $t_z\to t_{\mathrm{ret}}+z\rho'$; at fixed $t$ and $\rho$ of the field point, $t_{\mathrm{ret}}\to t-\rho$. Then
(\ref{eq:I-dS-tret})--(\ref{eq:J-dS-tret}) become
\begin{align}
I_L&=\int d^3x'\int_{-1}^{1}dz\;\Big\{\delta_\ell(z)\,\widehat{x}{\,}'_L\,\widehat{T}
+A_\ell\,\delta_{\ell+1}(z)\,\widehat{x}{\,}'_{aL}\,\partial_tT_{0a}
+B_\ell\,\delta_{\ell+2}(z)\,\widehat{x}{\,}'_{abL}\,\partial_t^2T_{ab}\Big\},
\label{eq:I-flat}\\[6pt]
J_L&=-\mathrm{STF}_L\int d^3x'\int_{-1}^{1}dz\;\Big\{\delta_\ell(z)\,\epsilon_{i_\ell ab}\,x'_a\,\widehat{x}{\,}'_{L-1}\,T_{0b}
+C_\ell\,\delta_{\ell+1}(z)\,\epsilon_{i_\ell ab}\,\widehat{x}{\,}'_{acL-1}\,\partial_tT_{bc}\Big\},
\label{eq:J-flat}
\end{align}
where the quantities in the brackets are evaluated at $\{t_{\mathrm{ret}}+z\rho',\vec{x}\,'\}$. Equations (\ref{eq:I-flat})--(\ref{eq:J-flat}) are \eqref{eq:I-L-main-tex}--\eqref{eq:J-L-main-tex} with $\Theta_{\mu\nu}\to T_{\mu\nu}$ and $\eta\to t$: the flat-space
Damour--Iyer moments of a conserved $T_{\mu\nu}$, with $\sigma=T^{00}+T^{kk}$, $\sigma_a=T^{0a}$, $\sigma_{ab}=T^{ab}$.

  The metric goes the same way. Since $t$ depends on $\eta$ alone, $\vx$ and the operators
$\pd_i$ are common to the two charts and $\dd\eta=\dd t/a$. The background metric becomes
\be
\dd\bar{s}^2=-\dd t^2+a^2\dd\vx^{\,2},
\label{eq:background-cosmic}
\ee
while the components of the perturbation transform as
\be
h_{tt}=a^{-2}h_{\eta\eta},\qquad h_{ti}=a^{-1}h_{\eta i},
\label{eq:transf}
\ee
with $h_{ij}$ unchanged.  With these inputs, the canonical metric
\eqref{eq:h-etaeta}--\eqref{eq:h-ij} can be readily transformed into the new coordinates. In these
coordinates the $H\to0$ limit is completely regular. What survives is the flat-space Damour--Iyer canonical metric, with all moments evaluated at the
flat retarded time $t-\rho$.

\section{The quadrupolar metric in canonical form}
\label{sec:comparison}

The exterior solution of Section~\ref{sec:metric} holds at arbitrary multipole order. We now
specialise it to the quadrupolar order and compare with the
quadrupolar metric of Comp\`ere, Hoque, and Kutluk (CHK) \cite{Compere:2023ktn}. The comparison is more than a check. In our language the quadrupolar order, namely the $\ell = 2$ canonical field, contains only the mass and the current quadrupole. Everything else is a residual gauge transformation, and, as we noted above, the pieces that enter the residual gauge transformation contain history integrals. As a result, it is not at all obvious how our results are related to those of Comp\`ere, Hoque, and Kutluk. Two things must be fixed before a comparison can be attempted. 
\begin{enumerate}
\item The first is the truncation. The moments \eqref{eq:I-dS}--\eqref{eq:J-dS} are organised by the STF rank $\ell$, while the quadrupolar truncation of  CHK is organised using Cartesian (non-STF) instantaneous source moments \eqref{eq:intro-instantaneous-moments}. The two are not the same truncation. 
\item The second is the gauge. Our exterior field is in the canonical form \eqref{eq:h-etaeta}--\eqref{eq:h-ij}, whereas
Comp\`ere, Hoque, and Kutluk work in the gauge in which the retarded solution was generated. The two differ by a residual de Sitter transformation as explained in
Section~\ref{sec:eom}. Its
generator is built from the four families $W_L$, $X_L$, $Y_L$, $Z_L$ of \eqref{eq:W-dS}--\eqref{eq:Z-dS}, evaluated in
the truncation. 
\end{enumerate}
To establish the equivalence we proceed in three steps. Section~\ref{sec:trunc} sets up the CHK truncation and the identities it implies.
Section~\ref{sec:can} gives the canonical field at quadrupolar order in the Cartesian moments. Section~\ref{sec:xi} constructs $\xi^\mu$ and establishes that the two presentations differ by a residual de Sitter transformation.

\subsection{Quadrupolar truncation}
\label{sec:trunc}
The source moments of CHK are the Cartesian moments \eqref{eq:intro-instantaneous-moments}. They are
integrals over a slice of constant time in which the powers of $a$ convert the coordinate volume and the coordinates $x'_i$ into orthonormal-frame components \cite{Ashtekar:2015lxa, Date:2015kma}. We call the number of indices in $L$ the \emph{degree} of such a moment, and keep the letter
$\ell$ for the rank of the STF moments $I_L$ and $J_L$. The quadrupolar truncation of
\cite{Compere:2023ktn} is the assumption that
\be
Q^{(\rho)}_L=P_{i|L}=S_{ij|L}=0\qquad\text{for every moment of degree three or higher.}
\label{eq:trunc}
\ee
It leaves $Q^{(\rho)}$, $Q^{(\rho)}_i$, $Q^{(\rho)}_{ij}$ for the energy density, $P_i$, $P_{i|j}$, $P_{i|jk}$ for the
momentum density, and $S_{ij}$, $S_{ij|k}$, $S_{ij|kl}$ for the stresses. Following \cite{Compere:2023ktn} we write
$Q^{(p)}_L\equiv S_{kk|L}$ for the moments of the pressure and $Q^{(\rho+p)}_L\equiv Q^{(\rho)}_L+Q^{(p)}_L$. In this
section all Cartesian moments are functions of the retarded cosmic time $t_{\rm ret}$, and we use the shorthand $\pd_t\equiv\pd_{t_{\rm ret}}$.

Two remarks are in order here to clarify the physical import of the truncation \eqref{eq:trunc}. These points are implicit in the presentation of \cite{Compere:2023ktn}. They were highlighted in our recent paper \cite[version 2 to be precise]{Harsh:2026vmh}. First, equation \eqref{eq:trunc} is a condition on the source, not an approximation made in
the field. Once it is imposed nothing else is expanded, neither in $H$, nor in the distance to the source, nor in the
rate at which the moments vary. It holds exactly for a point-like source, whose stress tensor is a combination of
$\delta^3(\vec x\,)$ and its first two derivatives with time-dependent coefficients. For a source of finite physical
size it is an approximation, in which each additional degree costs one more power of the source size.
Second, it is not a slow-motion truncation. A post-Newtonian expansion orders terms by the velocities inside the
source, and that ordering mixes degrees. For instance, \eqref{eq:trunc} keeps the pressure quadrupole $Q^{(p)}_{ij}$
but discards the degree-four moment that gives the  retardation correction to the mass quadrupole, although for a
slowly moving source both are of relative order $v^2$. No such assumption entered Section~\ref{sec:poincare}. The truncation is made
here only in order to compare with \cite{Compere:2023ktn}.

Let us recall that the exact moments \eqref{eq:I-dS}--\eqref{eq:J-dS} are organised differently, by the STF rank $\ell$, and a moment of
rank $\ell$ contains Cartesian moments of every degree from $\ell$ upwards. There are two reasons. By
\eqref{eq:intro-z-slow-motion} the light-cone average is the value on the slice $\eta=\etaret$ plus corrections in which
every two time derivatives come with a factor $\rho'^2$, that is, with two more degrees. And the $A_\ell$ and $C_\ell$
terms carry one more factor of $x'$ than the leading term, the $B_\ell$ term two more. Conversely, a Cartesian moment
is reducible: $Q^{(\rho+p)}_{ij}$ contains the quadrupole $Q^{(\rho+p)}_{\langle ij\rangle}$ and also the trace
$Q^{(\rho+p)}_{kk}$, which is a monopole.

The truncation \eqref{eq:trunc} is therefore not the same thing as keeping $\ell\le2$ in the all-$\ell$ solution. It
does imply it, however, since all $I_L$ and $J_L$ with $\ell\ge3$ vanish and the sums in
\eqref{eq:h-etaeta}--\eqref{eq:h-ij} stop at $\ell=2$. It also simplifies the moments that remain. At $\ell=2$
only the leading terms of
\eqref{eq:I-dS}--\eqref{eq:J-dS} survive, with the light-cone average collapsed onto the slice $\eta=\etaret$,
\be
I_{ij}=\aR^{-2}\,Q^{(\rho+p)}_{\langle ij\rangle},\qquad
J_{ij}=-\tfrac12\,\aR^{-2}\big(\epsilon_{iab}\,P_{b|aj}+\epsilon_{jab}\,P_{b|ai}\big).
\label{eq:quadrupoles}
\ee
These are the mass quadrupole and the current quadrupole. The powers of $\aR$ appear because $I_L$ and $J_L$ are
moments in the comoving coordinates coming from the auxiliary flat problem, while \eqref{eq:intro-instantaneous-moments} are moments in
proper distance. As in flat space \cite{Damour:1990gj}, the mass quadrupole is sourced by $T_{00}+T_{kk}$, the energy
density plus the trace of the stresses, and the current quadrupole by the momentum density alone. Note that $\tfrac43\aR^2J_{ij}$
is the current quadrupole of \cite{Compere:2023ktn}. At $\ell=0$ and $\ell=1$ the subleading terms of
\eqref{eq:I-dS}--\eqref{eq:J-dS} contribute as well, and for the monopole so does the first correction in
\eqref{eq:intro-z-slow-motion}, which brings in the trace $Q^{(\rho+p)}_{kk}$. To simplify these moments we need the
conservation law.

Integrating \eqref{eq:dS-conservation1}--\eqref{eq:dS-conservation2} against the weights of
\eqref{eq:intro-instantaneous-moments} gives one energy law and one momentum law at each degree
\cite{Compere:2023ktn}. At degree zero, one and two they read
\begin{align}
\pd_tQ^{(\rho)}&=-H\,Q^{(p)}, & \pd_tP_i&=-H\,P_i,
\label{eq:balance0}\\
\pd_tQ^{(\rho)}_i&=H\big(Q^{(\rho)}_i-Q^{(p)}_i\big)-P_i, & \pd_tP_{i|j}&=-S_{ij},
\label{eq:balance1}\\
\pd_tQ^{(\rho)}_{ij}&=H\big(2Q^{(\rho)}_{ij}-Q^{(p)}_{ij}\big)-P_{i|j}-P_{j|i}, & \pd_tP_{i|jk}&=H\,P_{i|jk}-S_{ij|k}-S_{ik|j},
\label{eq:balance2}
\end{align}
and they hold for any source. The antisymmetric part of the momentum law in \eqref{eq:balance1} is
the conservation of the angular momentum $\epsilon_{iab}P_{a|b}$. Its symmetric part, combined with the energy law in
\eqref{eq:balance2}, gives the important relation
\cite{Ashtekar:2015lxa,Date:2015kma,Compere:2023ktn},
\be
S_{ij}=\tfrac12\,\pd_t\Big[\pd_tQ^{(\rho)}_{ij}-2H\,Q^{(\rho)}_{ij}+H\,Q^{(p)}_{ij}\Big].
\label{eq:virial}
\ee

The truncation also enters via the degree three equations (and it ends there). With \eqref{eq:trunc} every moment of degree three drops out of the
degree three equations, and what is left are \emph{algebraic constraints on the moments of degree one and two} \cite{Compere:2023ktn},
\be
P_{(i|jk)}=0,\qquad S_{i(j|kl)}=0,\qquad S_{(ij|k)}=0,
\label{eq:constraints}
\ee
where the parentheses denote symmetrisation over all the indices they enclose. With these constraints the truncated moments describe a
conserved source in their own right, to which the results of Section~\ref{sec:poincare} apply without approximation.\footnote{A truncation is consistent only if it is compatible with the
conservation equations. The CHK truncation is consistent. For further discussion on this issue see \cite{Date:2024peu, Date:2026rft}.}
The contractions that we shall need are
\be
P_{a|ai}=-\tfrac12\,P_{i|kk},\qquad S_{ia|ja}=-\tfrac12\,Q^{(p)}_{ij},\qquad Q^{(p)}_i=\big(\pd_t-H\big)P_{i|kk}.
\label{eq:contractions}
\ee

With \eqref{eq:balance0}--\eqref{eq:contractions} the moments with $\ell\le1$ collapse to
\be
I=Q^{(\rho)}-H\,P_{a|a},\qquad
\aR\,I_i=Q^{(\rho)}_i+H\,P_{i|kk},\qquad
\aR\,J_i=\epsilon_{iab}\,P_{a|b}.
\label{eq:low-moments}
\ee
None of them is radiative. By \eqref{eq:balance0}--\eqref{eq:balance1}, $I$ is constant
\be
\pd_t I=-HQ^{(p)}-H\pd_tP_{a|a}=0;
\ee
it is the de Sitter energy. Likewise $\aR J_i$ is
the conserved angular momentum. The mass dipole $I_i$ changes at a rate set by the momentum, $\dot I_i = \aR \partial_{t} I_i=-P_i$. Strictly, for the sources of
Section~\ref{sec:eom}, which die away toward the past horizon, all three vanish. We keep them formally, as
\cite{Compere:2023ktn} do.

\subsection{Canonical field at quadrupolar order}
\label{sec:can}

With the truncation \eqref{eq:trunc} the sums in the canonical metric \eqref{eq:h-etaeta}--\eqref{eq:h-ij} stop at
$\ell=2$, and the five moments that remain are given in the Cartesian moments  by
\eqref{eq:quadrupoles} and \eqref{eq:low-moments}. In this subsection we write out the resulting field $\widehat\chi^{\rm can}$, $\chi_{0i}^{\rm can}$ and $\chi_{ij}^{\rm can}$.  

To proceed it is convenient to define the proper distance radial variable $r$
\be
r = a(\eta) \rho \implies
\frac1r = \frac{1}{\aR \rho}- H.
\ee
Moreover, $I_L$ and $J_L$ as defined  in \eqref{eq:I-dS}--\eqref{eq:J-dS} are moments in the comoving coordinates.  The corresponding moments in proper distance may be defined as $\bar I_L\equiv\aR^{\ell}I_L$ and $\bar J_L\equiv\aR^{\ell}J_L$. By \eqref{eq:quadrupoles} and \eqref{eq:low-moments} they are
\begin{align}
&\bar I_{ij}=Q^{(\rho+p)}_{\langle ij\rangle}, &
&\bar J_{ij}=-\tfrac12\big(\epsilon_{iab}\,P_{b|aj}+\epsilon_{jab}\,P_{b|ai}\big),\\
&\bar I_i=Q^{(\rho)}_i+H\,P_{i|kk}, & & \bar J_i=\epsilon_{iab}\,P_{a|b},
\label{eq:proper-moments}
\end{align}
and $\bar I = I$. The terms with $\ell\le1$ are
\be
\widehat\chi^{\rm can}\Big|_{\ell\le1}= \frac{4G}{r} \left[\bar{I}+n_i\left(\pd_t\bar I_i+\frac{\bar I_i}{r}\right)\right],
\qquad
\chi^{\rm can}_{0i}\Big|_{\ell\le1}= \frac{4G}{r}\left[P_i+\frac{\epsilon_{iab}\,n_a\bar J_b}{2\,r}\right],
\label{eq:low-field}
\ee
and they do not contribute to $\chi^{\rm can}_{ij}$. The terms with $\ell\le1$ solve the field equations. The gauge condition \eqref{eq:gh-condition} holds if and only if $P_i=0$: for $P_i\neq0$ one finds $\mathcal C_0=0$ and $\mathcal C_i=-4GH^2\aR P_i$, which is constant by \eqref{eq:balance0}. For a source in our class $P_i=0$. See also the discussion in  \cite{Date:2026rft} on this point. The difference between the two solutions, our canonical solution and the CHK solution,  does not depend
on $\bar I$, $\bar I_i$ and $\bar J_i$, so that the terms \eqref{eq:low-field} play no role in the comparison. Hence, for the rest of this section this subtlety does not matter. Since \eqref{eq:low-field} solves the field equations and the gauge condition for constant $\bar I$, $\bar J_i$ and $(\pd_t-H)\bar I_i=0$, the field in canonical form applies, formally, also to the sources of \cite{Compere:2023ktn} that carry charges, although the derivation does not.

The quadrupolar $(\ell=2)$  part of the canonical field is
\begin{align}
\widehat\chi^{\rm can}&=\frac{2G}{r}\,n_in_j\,\mathcal D\,\bar I_{ij},
\label{eq:can-hat}\\
\chi^{\rm can}_{0i}&=-\frac{2G}{r}\,n_j\Big(\pd_t+\frac{1}{r}\Big)\big(\pd_t-2H\big)\bar I_{ij}
+\frac{4G}{3r}\,\epsilon_{iab}\,n_an_j\,\mathcal D\,\bar J_{bj},
\label{eq:can-vec}\\
\chi^{\rm can}_{ij}&=\frac{2G}{r}\left(\pd_t + H^2 r\right)\left(\pd_t-2H\right)\bar I_{ij}
-\frac{8G}{3r}\,n_a\,\epsilon_{ab(i}\left(\pd_t+\frac{1}{r}+ H^2r\right)\big(\pd_t-H\big)\bar J_{j)b},
\label{eq:can-ten}
\end{align}
where
\be
\mathcal D\equiv \pd_t^2-H^2+\frac{3}{r} \pd_t+\frac{3}{r^2}.
\label{eq:calD}
\ee
The moments are evaluated at $t_{\rm ret}$. One checks directly that they satisfy the field equations
\eqref{eq:scalarChiEOM}--\eqref{eq:tensorChiEOM} outside the source and the gauge condition \eqref{eq:gh-condition},
for arbitrary functions $\bar I_{ij}(t_{\rm ret})$ and $\bar J_{ij}(t_{\rm ret})$. The field is local in the retarded time. As $r\to\infty$ at fixed $t_{\rm ret}$, that is, toward $\mathscr I^+$, $\widehat\chi^{\rm can}$ and
$\chi^{\rm can}_{0i}$ vanish, while
\be
\chi^{\rm can}_{ij}\;\to\;\chi^{(0)}_{ij}\equiv
2GH^2\left[\left(\pd_t-2H\right)\bar I_{ij}-\frac43\,n_a\,\epsilon_{ab(i}\left(\pd_t-H\right)\bar J_{j)b}\right].
\label{eq:chi0}
\ee
A field that survives on $\mathscr I^+$ is a characteristic feature of de Sitter radiation. The $1/r$ part of \eqref{eq:can-ten} is $H^{-2}\pd_t\chi^{(0)}_{ij}/r$, the same relation
between the two as in the solutions of \cite{Compere:2023ktn, Harsh:2026vmh}.

Equations \eqref{eq:low-field} and \eqref{eq:can-hat}--\eqref{eq:can-ten}  are the quadrupolar field in canonical form.
The solution of \cite{Compere:2023ktn} is the retarded solution of \eqref{eq:scalarChiEOM}--\eqref{eq:tensorChiEOM}
for the same source, in the same gauge. However, it depends on additional functions. In Section~\ref{sec:xi} we show
that this additional dependence is a residual gauge transformation.

\subsection{Residual gauge transformation}
\label{sec:xi}

The solution of \cite{Compere:2023ktn} is the retarded solution of
\eqref{eq:scalarChiEOM}--\eqref{eq:tensorChiEOM}. In our construction the retarded solution differs from its canonical form by a residual gauge transformation. The residual
de Sitter vector $\xi^\mu$ takes the form
\eqref{eq:w-to-xi-1}--\eqref{eq:w-to-xi-2}: \be
\xi^0=-a^{-1}w_0,\qquad \xi^i=a^{-1}w_i+H\int_{-\infty}^{\eta}d\eta'\,w_i(\eta',\vec x\,).
\label{eq:xi-map}
\ee
The vector $w_\mu$ is given by \eqref{eq:w0}--\eqref{eq:wi} in terms of the four families of moments of the physical stress tensor
\eqref{eq:W-dS}--\eqref{eq:Z-dS}.
We now show that
\be
\chi_{\mu\nu}^{\mathrm{CHK}}=\chi^{\rm can}_{\mu\nu}+\delta_\xi\chi_{\mu\nu},
\label{eq:xi-identity}
\ee
with $\delta_\xi\chi_{\mu\nu}$ given by \eqref{eq:dS-gauge-action}. On the left stands the solution restricted to our class. In the truncation \eqref{eq:trunc} every step of
this chain can be made explicit.

In the truncation only $W$, $W_i$, $X$, $Y_i$, $Y_{ij}$ and $Z_i$ contain moments of
degree two or less, and $Z_i$ vanishes by \eqref{eq:contractions} and the symmetry of $Q^{(p)}_{ij}$ in the indices $ij$. The others are
\begin{align}
\label{eq:families-1}
&W=\aR^{-1}\left[-\frac13P_{a|a}+\frac1{24}\left(\pd_t-3H\right)Q^{(p)}_{kk}\right],\\
&W_i=\frac14\,\aR^{-2}P_{i|kk},\\
&X=-\frac1{24}\,\aR^{-2} Q^{(p)}_{kk} + \frac{1}{24} H \!\int_{-\infty}^{t_{\rm ret}}\!dt'\,a(t')^{-2} Q^{(p)}_{kk}(t'),\\
&Y_i=-\frac14\,\aR^{-1}\left(\pd_t-2H\right)P_{i|kk},\\
&Y_{ij}=-\,\aR^{-2} Q^{(p)}_{\langle ij \rangle} + H \!\int_{-\infty}^{t_{\rm ret}}\!dt'\,a(t')^{-2}Q^{(p)}_{\langle ij \rangle}(t').
\label{eq:families-2}
\end{align}
Note that $X$ and $Y_{ij}$ contain an integral over the entire history of the degree-two pressure moment $Q^{(p)}_{ij}$. So the vector $w_i$ depends on
the whole past of the source.

The de Sitter vector does not. Inserting \eqref{eq:families-1}--\eqref{eq:families-2} in \eqref{eq:w0}--\eqref{eq:wi} and then in
\eqref{eq:xi-map}, we find
\begin{align}
\label{eq:xi-1}
\xi^0&=-\frac{G}{r \aR}\left[-\frac43P_{a|a}+\frac16\left(\pd_t-3H\right)Q^{(p)}_{kk}
+n_i\left(\pd_t-H+\frac1r\right)P_{i|kk}\right],\\
\xi^i&=\frac{G}{r \aR}\left[\left(\pd_t-H+ H^2 r\right)P_{i|kk}
+\left(\pd_t-H+H^2r+\frac{1}{r}\right)\left(2\,n_jQ^{(p)}_{ij}-\frac12\,n_iQ^{(p)}_{kk}\right)\right].
\label{eq:xi-2}
\end{align}
The history integrals have cancelled between the two terms of $\xi^i$ in \eqref{eq:xi-map}. 
With \eqref{eq:xi-1}--\eqref{eq:xi-2} the identity \eqref{eq:xi-identity} can be verified directly. The moments with $\ell\le1$ do not enter $\xi^\mu$, whatever their values. 
One also checks
that $\xi^\mu$ satisfies the conditions of Section~\ref{sec:eom} for a residual transformation.

\section{Conclusion}
\label{sec:conclusion}

In this paper, we have derived the linearised gravitational waves generated by a spatially compact source in the Poincar\'e patch of de Sitter spacetime at arbitrary multipole order. The construction rests on two observations. First, an integral of the spatial stresses over the past converts the physical stress tensor $T_{\mu\nu}$ into an auxiliary stress tensor $\Theta_{\mu\nu}$ that is conserved with respect to the auxiliary Minkowski metric. Second, the transformation $\chi_{ij}=\eta D_-\Xi_{ij}$ removes the curvature potential from the tensor equation, so that all three sectors obey flat wave equations. Together with the correspondence between residual gauge freedoms, these observations reduce the de Sitter problem to the flat-space linearised problem in harmonic gauge.

The flat-space multipole results of Damour and Iyer then allow us to obtain our main results, namely,  the exterior metric perturbation at arbitrary $\ell$, Eqs.~\eqref{eq:h-etaeta}--\eqref{eq:h-ij}, and the irreducible moments as light-cone integrals of the physical stress tensor, Eqs.~\eqref{eq:I-dS}--\eqref{eq:J-dS}. The organisation into irreducible moments from the outset makes the analysis tractable at arbitrary $\ell$.  The de Sitter tail is carried by the auxiliary stress tensor. The moments $I_L$, $J_L$
and the canonical metric are local in the retarded time.  In cosmic time coordinates, the limit $H\to0$ is regular and returns the flat-space Damour--Iyer moments and canonical metric.

As a check, Section~\ref{sec:comparison} recovers the quadrupolar metric of Comp\`ere, Hoque, and Kutluk
\cite{Compere:2023ktn}. In their truncation, it is our canonical field ~\eqref{eq:h-etaeta}--\eqref{eq:h-ij} together with a residual gauge
transformation, which we constructed explicitly. Appendix~\ref{app:retarded} compares the retarded solution itself, truncated at an arbitrary degree, directly with \cite{Compere:2023ktn} and with the octupolar metric of \cite{Harsh:2026vmh}; no gauge transformation is involved there. We have not carried out a comparison of the canonical field with the octupolar metric through a residual gauge transformation. We do not foresee any difficulty of principle in doing so. The computation is tedious, and for this reason alone we leave it for the future.

Our focus here has been on the construction of the field and on its mapping to
flat spacetime. One consequence of the construction deserves to be recorded,
because it is the starting point for what comes next. On $\mathscr I^+$, that
is, as $\eta\to0$ at fixed $\vec x$, the scalar and vector parts of the
canonical field vanish, while by \eqref{eq:chi-ij} the tensor part has the
finite limit
$\chi^{\rm can}_{ij}\big|_{\mathscr I^+}=H\,Q^{\rm can}_{ij}\big|_{\eta=0}$. This is the field on $\mathscr I^+$ at every
multipole order, in terms of the moments $I_L$ and $J_L$ alone. The energy flux through $\mathscr I^+$ and the displacement memory are
functionals of the field on $\mathscr I^+$
\cite{Ashtekar:2015lxa,Compere:2023ktn}, and have so far been obtained at low
multipole orders. With the expression above they can be studied at all
multipole orders, together with the question of how the residual gauge
transformations of Section~\ref{sec:gauge} act on them. We will take this up
in a forthcoming paper. In that paper we will also give the flux-balance laws \cite{Compere:2024ekl} associated with
the dilatation, that is, the time translation of de Sitter, and with the rotations, in terms of
the moments $I_L$ and $J_L$. For the computation we plan to use the auxiliary flat problem once
more.

Shortly after this work was completed, the preprint by Comp\`ere and
Hoque \cite{Compere:2026xwy} appeared. It treats compact sources on
spatially flat FLRW backgrounds with general expansion histories,
obtaining the retarded linearised field at every multipole order,
including the tails, together with the disturbance induced in the
cosmic fluid --- a response that is absent in pure de Sitter
spacetime. Specialised to de Sitter, their results partially overlap
with ours; in particular, their construction reproduces the
quadrupolar and octupolar models of
\cite{Compere:2023ktn,Harsh:2026vmh}. The two constructions differ in method: theirs works directly with the retarded Green functions of the curved spacetime, whereas ours proceeds through an exact map to an auxiliary problem in flat spacetime. A detailed comparison of the two at arbitrary multipole order would be worthwhile and is left for future work.

\bigskip

\paragraph{Acknowledgements:} 
We thank K.~G.~Arun, Jahanur Hoque, R.~Loganayagam and George Zahariade for helpful conversations, and are especially grateful to Ghanashyam Date for numerous discussions and for his comments on an earlier draft. The work of A.V.\ was partly supported by SERB Core Research Grant CRG/2023/000545.  O.S.\ is supported by SONATA BIS grant 2023/50/E/ST2/00231 from the Polish National Science Centre and acknowledges his debt to the people of Poland for their steady and generous support of research in basic sciences.  We also acknowledge the support from the Infosys Foundation to CMI.

\bigskip

\paragraph{Statement on AI use:} 

In the course of this work we used Claude Fable 5.1 (Anthropic) and Gemini 3.1 Pro (Google) to discuss ideas, to assist with calculations, many of them exploratory and not part of the final paper, and to check drafts of the manuscript. Claude also helped with preliminary drafts of parts of the text. These drafts were critically reviewed and revised by us, and the final text is our own. All results presented here were independently verified by the authors, who take full responsibility for the content of this paper.

\appendix
\section{Irreducible Cartesian multipoles and the Damour--Iyer results}
\label{app:stf}
This appendix fixes the symmetric trace-free (STF) conventions used in the main text and states the flat-space results of Damour and Iyer (DI)~\cite{Damour:1990gj}. The notation from the main text is as follows: field point $\vec x$, $\rho=|\vec x|$, $n_i=x_i/\rho$; source point $\vec x{\,}'$, $\rho'=|\vec x{\,}'|$, $n'_i=x'_i/\rho'$; retarded time $\etaret=\eta-\rho$; light-cone parameter $z\in[-1,1]$ and $\eta_z=\etaret+z\rho'$.

A capital letter denotes a multi-index and the corresponding lower-case letter its length: $L=i_1i_2\cdots i_\ell$, $L-1=i_1\cdots i_{\ell-1}$, $aL=a\,i_1\cdots i_\ell$, and
\begin{equation}
x_L=x_{i_1}x_{i_2}\cdots x_{i_\ell},\qquad
\pd_L=\pd_{i_1}\pd_{i_2}\cdots\pd_{i_\ell},\qquad
n_L=n_{i_1}\cdots n_{i_\ell}.
\label{eq:multiindex}
\end{equation}
Parentheses denote symmetrisation with weight $1/\ell!$, angle brackets denote the symmetric trace-free (STF) part, and a hat is short-hand for the STF:
\begin{equation}
T_{\langle L\rangle}\equiv\STF_L(T_L) \equiv \widehat{T}_L,\qquad
\widehat{x}_L\equiv x_{\langle L\rangle},\qquad
\widehat{n}_L\equiv n_{\langle L\rangle}=\rho^{-\ell}\widehat{x}_L .
\label{eq:hat}
\end{equation}
The STF part of a rank-$\ell$ tensor is given explicitly by \cite{Blanchet:1985sp, Damour:1990gj}
\begin{align}
\label{eq:stf-projector}
& T_{\langle L\rangle}=\sum_{k=0}^{[\ell/2]}a_{\ell k}\;
\delta_{(i_1i_2}\cdots\delta_{i_{2k-1}i_{2k}}\,S_{i_{2k+1}\cdots i_\ell)\,a_1a_1\cdots a_ka_k},\\
& S_L = T_{(L)}, \\
& a_{\ell k}=(-1)^k\,\frac{\ell!\,(2\ell-2k-1)!!}{(\ell-2k)!\,(2\ell-1)!!\,(2k)!!}.
\end{align}
For $\ell=2,3$,
\begin{equation}
T_{\langle ij\rangle}=T_{(ij)}-\tfrac13\delta_{ij}T_{ss},\qquad
T_{\langle ijk\rangle}=T_{(ijk)}-\tfrac15\big(\delta_{ij}T_{(kss)}+\delta_{jk}T_{(iss)}+\delta_{ki}T_{(jss)}\big).
\label{eq:stf-23}
\end{equation}
To move a fixed index in or out of an STF projection one uses the peeling formula (DI Eq.~(2.14)), valid for any vector $V_a$ and STF $\widehat T_L$,
\begin{equation}
(\ell+1)\,V_{\langle a} \widehat T_{L\rangle}
=V_a \widehat T_L+\ell\, \widehat T_{a\langle L-1}V_{i_\ell\rangle}-\frac{2\ell}{2\ell+1}\,V_b \widehat T_{b\langle L-1}\delta_{i_\ell\rangle a}.
\label{eq:peeling}
\end{equation}
The analogous six-term decomposition of a symmetric tensor times an STF tensor is not reproduced here. The manipulations in DI use that decomposition in an important way, but we do not need those details here. Only DI's end results are used in this paper. 

For any STF tensor function $ F_L(\etaret)$ of the retarded time, the function $\phi = \pd_L[F_L(\etaret)/\rho]$ solves $\Box\phi=0$ for $\rho>0$. Its explicit form is
\begin{equation}
\pd_L\!\left[\frac{F_L(\etaret)}{\rho}\right]
=(-1)^\ell\sum_{k=0}^{\ell}\frac{(\ell+k)!}{2^k\,k!\,(\ell-k)!}\;
\frac{\widehat{n}_L\,F_L^{(\ell-k)}(\etaret)}{\rho^{k+1}},
\label{eq:multipole-wave}
\end{equation}
so that the $1/\rho$ (radiative) term carries $\ell$ time derivatives, the $1/\rho^{\ell+1}$ (static) term carries none.\footnote{Identity \eqref{eq:multipole-wave} follows from the  fact that for any radial function $f(\rho)$,
\begin{equation}
\widehat\partial_Lf (\rho)=\widehat n_L \rho^\ell\left(\frac1\rho\frac{d}{d\rho}\right)^{\!\ell}f (\rho).
\end{equation} Expanding the radial derivatives gives the identity.} Equation~\eqref{eq:multipole-wave} is the only formula needed to convert the exterior solutions into explicit functions of $(\eta,\vec x)$. Let $S(\eta,\vec x{\,}')$ vanish for $\rho'>R_s$ and let $V$ be the retarded solution of $\Box V=-4\pi S$,
\begin{equation}
V(\eta,\vec x)=\int\dd^3x'\;\frac{S\big(\eta-|\vec x-\vec x{\,}'|,\vec x{\,}'\big)}{|\vec x-\vec x{\,}'|}.
\label{eq:scalar-retarded}
\end{equation}
Then, for $\rho>R_s$ (DI Eqs.~(3.3)--(3.6)),
\begin{equation}
V(\eta,\vec x)=\sum_{\ell\ge0}\frac{(-1)^\ell}{\ell!}\,\pd_L\!\left[\frac{F_L(\etaret)}{\rho}\right],
\qquad
F_L(\etaret)=\int\dd^3x'\;\widehat{x}{}'_L\int_{-1}^{1}\dd z\;\delta_\ell(z)\,S(\eta_z,\vec x{\,}'),
\label{eq:scalar-theorem}
\end{equation}
with $\eta_z=\etaret+z\rho'$ and $\delta_\ell$ given as
\begin{equation}
\delta_\ell(z)=\frac{(2\ell+1)!!}{2^{\ell+1}\,\ell!}\,(1-z^2)^\ell,
\qquad
\int_{-1}^{1}\delta_\ell(z)\,\dd z=1.
\label{eq:delta-def}
\end{equation}
No expansion in the size or the internal velocity of the source is made. Equation \eqref{eq:scalar-theorem} is exact and holds in the whole exterior, near zone included. The $z$-integral has a causal meaning. A source point at radius $\rho'$ is seen from the field point at the time $\eta-|\vec x-\vec x{\,}'|$, and $\rho-\rho'\le|\vec x-\vec x{\,}'|\le\rho+\rho'$, so that time lies in $[\etaret-\rho',\etaret+\rho']$. The weight $\delta_\ell(z)$ is even, so only even moments are non-zero:
\begin{equation}
\int_{-1}^{1}\dd z\;\delta_\ell(z)\,z^{2p}=\frac{(2\ell+1)!!\,(2p-1)!!}{(2\ell+2p+1)!!},
\qquad
\int_{-1}^{1}\dd z\;\delta_\ell(z)\,z^{2p+1}=0 .
\label{eq:delta-moments}
\end{equation}
Taylor-expanding the source in $z\rho'$ under the integral and using \eqref{eq:delta-moments} gives the slow-motion form of the average,
\begin{align}
\int_{-1}^{1}\dd z\,& \delta_\ell(z)\,S(\etaret+z\rho', \vec x{\,}')
=\sum_{p\ge0}\frac{(2\ell+1)!!}{(2p)!!\,(2\ell+2p+1)!!}\,\rho'{}^{\,2p}S^{(2p)}(\etaret, \vec x{\,}')\\
&=S(\etaret)+\frac{\rho'{}^{\,2}}{2(2\ell+3)}S^{(2)}(\etaret, \vec x{\,}')+\frac{\rho'{}^{\,4}}{8(2\ell+3)(2\ell+5)}S^{(4)}(\etaret, \vec x{\,}')+\cdots,
\label{eq:delta-slow}
\end{align}
where $S^{(n)}(\etaret, \vec x{\,}') = (\partial^n/\partial \etaret^n) S(\etaret, \vec x{\,}').$
The series \eqref{eq:delta-slow} is the formal expansion of a hypergeometric operator,
\begin{equation}
\int_{-1}^{1}\dd z\,\delta_\ell(z)\,S(\etaret+z\rho', \vec x{\,}')
={}_0F_1\!\Big(\ell+\tfrac32;\tfrac14\rho'^{\,2}\pd_{\etaret}^2\Big)S(\etaret, \vec x{\,}'),
\label{eq:delta-resum}
\end{equation}
which allows us to write, 
\be
\label{eq:F_L-final}
F_L(\etaret)=\int\dd^3x'\;\widehat{x}{\,}'_L\;{}_0F_1\!\Big(\ell+\tfrac32;\tfrac14\rho'^{\,2}\pd_{\etaret}^2\Big)S(\etaret, \vec x{\,}').
\ee
This is the form used in Section \ref{sec:poincare}.

Let $\Theta_{\mu\nu}(\eta,\vec x)$ be a conserved symmetric tensor  with respect to the auxiliary Minkowski metric. Moreover, assume that $\Theta_{\mu\nu}(\eta,\vec x)$ vanishes for $\rho > R_s$ for all $\eta$.  Let $\mathfrak h_{\mu\nu}$ be the retarded solution of $\Box\mathfrak h_{\mu\nu}=-16\pi G\,\Theta_{\mu\nu}$ so that $\pd^\mu\mathfrak h_{\mu\nu}=0$.  Applying \eqref{eq:scalar-theorem} to each component, we have
\bea
&&
\mathfrak h^{\mu\nu}(\eta,\vec x)=4G\sum_{\ell\ge0}\frac{(-1)^\ell}{\ell!}\,\pd_L\!\left[\frac{F^{\mu\nu}_L(\etaret)}{\rho}\right],
\label{eq:raw-expansion} \\
&& F^{\mu\nu}_L(\etaret)=\int\dd^3x'\;\widehat{x}{\,}'_L\int_{-1}^{1}\dd z\;\delta_\ell(z)\,\Theta^{\mu\nu}(\eta_z,\vec x{\,}'). \label{eq:raw-expansion2}
\eea
The tensors $F^{0i}_L$ and $F^{ij}_L$ are STF only in $L$ and are therefore reducible. The content of DI Sec.~V is that conservation and the residual harmonic gauge freedom reduce the ten independent STF families $F^{\mu\nu}_L$ to two. For $\ell\ge0$ and $\ell\ge1$ respectively define (DI Eqs.~(5.33), (5.35))
\begin{align}
I_L(\etaret)&= \int\dd^3x'\int_{-1}^{1}\dd z\;\Big\{
\delta_\ell\,\widehat{x}{\,}'_L\,\widetilde\sigma
-A_\ell\,\delta_{\ell+1}\,\widehat{x}{\,}'_{aL}\,\pd_{\etaret}\widetilde\sigma_a
+B_\ell\,\delta_{\ell+2}\,\widehat{x}{\,}'_{abL}\,\pd^2_{\etaret}\widetilde\sigma_{ab}\Big\},
\label{eq:I-L}\\
J_L(\etaret)&=\STF_L\int\dd^3x'\int_{-1}^{1}\dd z\;\Big\{
\delta_\ell\,\epsilon_{i_\ell ab}\,x'_a\,\widehat{x}{\,}'_{L-1}\,\widetilde\sigma_b
-C_\ell\,\delta_{\ell+1}\,\epsilon_{i_\ell ab}\,\widehat{x}{\,}'_{acL-1}\,\pd_{\etaret}\widetilde\sigma_{bc}\Big\},
\label{eq:J-L}
\end{align}
with
\begin{align}
\sigma &= \Theta_{00}+ \Theta_{kk}, &
\sigma_a &= -\Theta_{0a}, &
\sigma_{ab} &= \Theta_{ab},
\end{align}
and tilde denotes evaluation at the light-cone time of the source point
\begin{align}
\widetilde\sigma &\equiv\sigma(\eta_z,\vec x{\,}'),&
\widetilde\sigma_a&\equiv\sigma_a(\eta_z,\vec x{\,}'),&
\widetilde\sigma_{ab}&\equiv\sigma_{ab}(\eta_z,\vec x{\,}'),&
\eta_z=\etaret+z\rho',
\label{eq:tilde}
\end{align}
together with
\begin{equation}
A_\ell=\frac{4(2\ell+1)}{(\ell+1)(2\ell+3)},\qquad
B_\ell=\frac{2(2\ell+1)}{(\ell+1)(\ell+2)(2\ell+5)},\qquad
C_\ell=\frac{2\ell+1}{(\ell+2)(2\ell+3)}.
\label{eq:ABC}
\end{equation}

The main result of DI states that modulo a residual gauge transformation, the exterior field is determined by $I_L$ and $J_L$ alone:
\begin{equation}
\mathfrak h_{\mu\nu}=\mathfrak h^{\rm can}_{\mu\nu}+\pd_\mu w_\nu+\pd_\nu w_\mu-\eta_{\mu\nu}\,\pd_\alpha w^\alpha,
\qquad \Box w_\mu=0\quad(\rho>R_s),
\label{eq:theorem}
\end{equation}
where the canonical field is
\begin{align}
\mathfrak h^{\rm can}_{00}&=4G\sum_{\ell\ge0}\frac{(-1)^\ell}{\ell!}\;\pd_L\!\left[\frac{I_L(\etaret)}{\rho}\right],
\label{eq:can-00}\\
\mathfrak h^{\rm can}_{0i}&=4G\sum_{\ell\ge1}\frac{(-1)^\ell}{\ell!}\left\{
\pd_{L-1}\!\left[\frac{\dot I_{iL-1}(\etaret)}{\rho}\right]
+\frac{\ell}{\ell+1}\,\epsilon_{iab}\,\pd_{aL-1}\!\left[\frac{J_{bL-1}(\etaret)}{\rho}\right]\right\},
\label{eq:can-0i}\\
\mathfrak h^{\rm can}_{ij}&=4G\sum_{\ell\ge2}\frac{(-1)^\ell}{\ell!}\left\{
\pd_{L-2}\!\left[\frac{\ddot I_{ijL-2}(\etaret)}{\rho}\right]
+\frac{2\ell}{\ell+1}\,\pd_{aL-2}\!\left[\frac{\epsilon_{ab(i}\dot J_{j)bL-2}(\etaret)}{\rho}\right]\right\}.
\label{eq:can-ij}
\end{align}
A dot  on a function of $\etaret$ denotes $\dd/\dd\etaret$.

The gauge vector $w_\mu$ is built from four further STF families  $W_L,X_L,Y_L,Z_L$:
\begin{align}
w_0&=4G\sum_{\ell\ge0}\frac{(-1)^\ell}{\ell!}\,\partial_L\!\left[\frac{W_L(\etaret)}{\rho}\right],
\label{eq:w0}\\
w_i&=4G\sum_{\ell\ge0}\frac{(-1)^\ell}{\ell!}\,\partial_{iL}\!\left[\frac{X_L(\etaret)}{\rho}\right]
\nonumber\\
&\quad+4G\sum_{\ell\ge1}\frac{(-1)^\ell}{\ell!}\left\{\partial_{L-1}\!\left[\frac{Y_{iL-1}(\etaret)}{\rho}\right]
+\frac{\ell}{\ell+1}\,\epsilon_{iab}\,\partial_{aL-1}\!\left[\frac{Z_{bL-1}(\etaret)}{\rho}\right]\right\}.
\label{eq:wi}
\end{align}
Each term is a multipole wave, so $\Box w_\mu=0$. $Y_L$ and $Z_L$ enter only for $\ell\ge1$ and $W_L$ and $X_L$ from
$\ell=0$. Explicit expressions are (see also \cite[Eq.~(141)]{Blanchet:2013haa}):
\begin{align}
W_L&=\int d^3x'\!\int_{-1}^{1}\!dz\,\Big\{\tfrac14A_\ell\,\delta_{\ell+1}\,\widehat{x}{\,}'_{aL}\,\widetilde\sigma_a
-\tfrac14B_\ell\,\delta_{\ell+2}\,\widehat{x}{\,}'_{abL}\,\pd_{\etaret} \widetilde\sigma_{ab}\Big\},
\label{eq:W-aux}\\
X_L&=\tfrac14B_\ell\int d^3x'\!\int_{-1}^{1}\!dz\;\delta_{\ell+2}\,\widehat{x}{\,}'_{abL}\,\widetilde\sigma_{ab},
\label{eq:X-aux}\\
Y_L&=\int d^3x'\!\int_{-1}^{1}\!dz\,\Big\{-\delta_\ell\,\widehat{x}{\,}'_L\,\widetilde\sigma_{aa}
+\tfrac34A_\ell\,\delta_{\ell+1}\,\widehat{x}{\,}'_{aL}\,\partial_{\etaret}\widetilde\sigma_a
-B_\ell\,\delta_{\ell+2}\,\widehat{x}{\,}'_{abL}\,\partial^2_{\etaret}\widetilde\sigma_{ab}\Big\},
\label{eq:Y-aux}\\
Z_L&=C_\ell\,\STF_L\int d^3x'\!\int_{-1}^{1}\!dz\;\delta_{\ell+1}\,\epsilon_{i_\ell ab}\,\widehat{x}{\,}'_{acL-1}\,\widetilde\sigma_{bc}.
\label{eq:Z-aux}
\end{align}

\section{The retarded solution in a truncation of arbitrary degree}
\label{app:retarded}

In Section~\ref{sec:comparison} the quadrupolar metric of \cite{Compere:2023ktn} was compared with the canonical
field. That comparison tests the main results of this paper, but it is indirect, since the two fields differ by a
residual gauge transformation. The metric of \cite{Compere:2023ktn} and its octupolar extension \cite{Harsh:2026vmh}
are retarded solutions of \eqref{eq:scalarChiEOM}--\eqref{eq:tensorChiEOM}, and in our construction the
retarded solution is available before any reduction to canonical form. Therefore, a direct comparison is possible.  In this appendix we evaluate the retarded solution in a truncation of the type used in
\cite{Compere:2023ktn, Harsh:2026vmh}, made at an arbitrary degree, and compare with those references term by term. No
gauge transformation is involved.

Every retarded integral in \eqref{eq:direct-scalar}--\eqref{eq:direct-Xi} is of the form
$\int d^3x'\,S(\eta-|\vec x-\vec x\,'|,\vec x\,')/|\vec x-\vec x\,'|$, with $S=\widehat T/\eta$, $T_{0i}/\eta$ or
$\eta\,\Pi_{ij}$. In \cite{Compere:2023ktn, Harsh:2026vmh} it is evaluated by expanding the integrand in the position
$\vec x\,'$ of the source point. The expansion is a Taylor series,
\be
\frac{S(\eta-|\vec x-\vec x\,'|,\vec x\,')}{|\vec x-\vec x\,'|}
=\sum_{q\ge0}\frac{(-1)^q}{q!}\;x'_L\,\pd_L\!\left[\frac{S(\etaret,\vec x\,')}{\rho}\right],\qquad |L|=q .
\label{eq:B-taylor}
\ee
After integration over the source, the term of degree $q$ is a Cartesian moment of $S$ of degree $q$, taken on the
slice $\eta'=\etaret$, contracted with $q$ derivatives of a spherical wave.

The expansion \eqref{eq:scalar-expansion} is organised differently, by the rank $\ell$ of STF tensors, and its
moments are averaged along the light cone. The two are related as follows. The product $x'_L$ is reducible. It
contains the STF tensors of rank $\ell=q-2k$, with $k=0,1,\dots,[q/2]$, each multiplied by $\rho'^{2k}$ and by $k$
Kronecker deltas~\cite[Eq.~A21a]{Blanchet:1985sp}. In \eqref{eq:B-taylor} a Kronecker delta contracts two
derivatives into a Laplacian, and the Laplacian acts on the spherical wave as $\pd_{\etaret}^2$, since
$F(\etaret)/\rho$ solves the wave equation. Collecting the terms of a given rank one finds
\be
\sum_{q\ge0}\frac{(-1)^q}{q!}\;x'_L\,\pd_L\!\left[\frac{S}{\rho}\right]
=\sum_{\ell\ge0}\frac{(-1)^\ell}{\ell!}\;\widehat x{\,}'_L\,\pd_L\!\left[\frac{1}{\rho}\sum_{k\ge0}
\frac{(2\ell+1)!!}{(2k)!!\,(2\ell+2k+1)!!}\;\rho'{}^{\,2k}\,\pd_{\etaret}^{2k}S\right],
\label{eq:B-byrank}
\ee
where $S$ stands for $S(\etaret,\vec x\,')$, $|L|=q$ on the left and $|L|=\ell$ on the right, and the terms with
$\ell+2k=q$ on the right make up the term of degree $q$ on the left. 
The expansion
\eqref{eq:B-byrank} is the exact expansion \eqref{eq:scalar-expansion} with the light cone average expanded in the size of the
source: the two are the same double series, summed in different orders. A Cartesian moment of degree $q$ feeds the
multipoles of rank $q,\,q-2,\dots$ through its traces. Conversely, a multipole of rank $\ell$ collects the traces of
the moments of degree $\ell,\,\ell+2,\dots$, and the light-cone average is their sum.

We call truncation at degree $N$ the assumption that
\be
Q^{(\rho)}_L=P_{i|L}=S_{ij|L}=0\qquad\text{for every moment of degree higher than $N$,}
\label{eq:B-trunc}
\ee
which is \eqref{eq:trunc} for $N=2$ \cite{Compere:2023ktn} and the octupolar truncation of \cite{Harsh:2026vmh} for
$N=3$. 
In such a truncation both series terminate. From \eqref{eq:B-byrank} only the
terms with $\ell+2k\le N$ survive under the integral sign, and the light-cone average collapses to a differential operator of finite order
acting on moments taken on the slice $\eta'=\etaret$. This is the precise sense in which the radiative moments of
Section~\ref{sec:poincare} reduce to the instantaneous moments \eqref{eq:intro-instantaneous-moments}.

We use the variables of Section~\ref{sec:comparison}: the retarded cosmic time $t_{\rm ret}$, the scale factor
$\aR=e^{Ht_{\rm ret}}$, the physical radius $r=a(\eta)\rho$, and $\pd_t\equiv\pd_{t_{\rm ret}}$. All Cartesian moments
are functions of $t_{\rm ret}$. The dictionary is
\be
\etaret=-\frac{1}{H\aR},\qquad \frac{1}{\aR\,\rho}=H+\frac1r,\qquad \frac{\eta}{\etaret}=\frac{\aR}{a}=\frac{1}{1+Hr},
\label{eq:B-dictionary}
\ee
and, for any function $X(t_{\rm ret})$ and integers $q,m\ge0$,
\be
\pd_{\etaret}^{\,m}\big[\aR^{-q}X\big]=\aR^{m-q}\,\prod_{s=0}^{m-1}\big(\pd_t-(q-s)H\big)X,
\label{eq:B-derivatives}
\ee
which is the moment differentiation identity of \cite{Harsh:2026vmh} in another form.

The sources of \eqref{eq:scalarChiEOM}--\eqref{eq:vectorChiEOM} are $\widehat T/\eta$ and $T_{0i}/\eta$. By
\eqref{eq:intro-instantaneous-moments} their moments of degree $q$ on the slice $\eta'=\etaret$ are
$-H\aR^{-q}Q^{(\rho+p)}_L$ and $-H\aR^{-q}P_{i|L}$, so that the terms of degree $q$ of
\eqref{eq:direct-scalar}--\eqref{eq:direct-vector} are
\bea
\label{eq:B-degree-q-1}
\widehat\chi^{[q]}&=&-4GH\eta\,\frac{(-1)^q}{q!}\,\pd_L\!\left[\frac{\aR^{-q}\,Q^{(\rho+p)}_L}{\rho}\right],\\
\chi^{[q]}_{0i}&=&-4GH\eta\,\frac{(-1)^q}{q!}\,\pd_L\!\left[\frac{\aR^{-q}\,P_{i|L}}{\rho}\right].
\label{eq:B-degree-q-2}
\eea
Each of them solves the homogeneous field equation outside the source, for arbitrary time dependence of the moment.

For the tensor sector we start from \eqref{eq:direct-Xi} and $\chi_{ij}=\eta D_-\Xi_{ij}$. Carrying out the derivative
under the integral, with $\pd_\eta\Pi_{ij}=T_{ij}/\eta^2$, gives
\be
\chi_{ij}=4G\!\int\! d^3x'\;\frac{\eta}{\eta-|\vec x-\vec x\,'|}\;
\frac{T_{ij}(\eta-|\vec x-\vec x\,'|,\vec x\,')}{|\vec x-\vec x\,'|}
\;+\;4G\!\int\! d^3x'\;\Pi_{ij}\big(\eta-|\vec x-\vec x\,'|,\vec x\,'\big),
\label{eq:B-sharp-tail}
\ee
which is \eqref{eq:direct-tensor} before the integration by parts that was made there. The two terms of \eqref{eq:B-sharp-tail} are the contributions of the
two parts of the retarded Green function of \eqref{eq:tensorChiEOM} \cite{Ashtekar:2015lxa,deVega:1998ia}: the first
comes from its support on the light cone, the second from its support inside. We call them the sharp part and the tail
part. Neither solves \eqref{eq:tensorChiEOM} by itself. Only the sum does.  The sharp part has the form of
\eqref{eq:direct-scalar}--\eqref{eq:direct-vector}. The tail part is 
$\Pi_{ij}$ of \eqref{eq:Pi-def}, integrated over the past light cone of the field point.

Two properties of the tail part follow directly. The first concerns $\mathscr I^+$. As $\eta\to0$ at fixed $\vec x$
the sharp part vanishes, like $\widehat\chi$ and $\chi_{0i}$, because of the factor $\eta$. Hence
\be
\chi_{ij}\big|_{\mathscr I^+}=4G\!\int\! d^3x'\;\Pi_{ij}\big(-|\vec x-\vec x\,'|,\vec x\,'\big).
\label{eq:B-scri}
\ee
The field that survives on $\mathscr I^+$ is entirely due to the tail.

The second property concerns locality in time. By \eqref{eq:Pi-def} and \eqref{eq:intro-instantaneous-moments} the moment of
degree $q$ of $\Pi_{ij}$ is $H^2\!\int^{t_{\rm ret}}\!dt'\,a^{-q}S_{ij|L}$, and the term of degree $q$ of the tail part is
\be
\tau^{[q]}_{ij}=4GH^2\,\frac{(-1)^q}{q!}\;\pd_L\!\left[\int_{-\infty}^{t_{\rm ret}}\!dt'\,a^{-q}(t')\,S_{ij|L}(t')\right].
\label{eq:B-tail-q}
\ee
There is no factor $1/\rho$: the bracket depends on $\vec x$ only through $\etaret$.  For $q\ge1$ the bracket is therefore always differentiated at least once with respect to $\etaret$,
and its first derivative is $\aR^{1-q}S_{ij|L}$, which is local. The history integral survives at $q=0$ only.
There the momentum law in \eqref{eq:balance1}, which holds for any source, gives
\be
\tau^{[0]}_{ij}=4GH^2\!\int_{-\infty}^{t_{\rm ret}}\!dt'\,S_{ij}(t')=-4GH^2\,P_{(i|j)}(t_{\rm ret})+C_{ij}(-\infty),
\label{eq:B-tail-0}
\ee
with $C_{ij}(-\infty)\equiv4GH^2\lim_{t\to-\infty}P_{(i|j)}$. Thus the retarded field is local in $t_{\rm ret}$ at every degree up to the constant $C_{ij}(-\infty)$, which vanishes for the sources of Section~2 since $P_{(i|j)}\to0$ toward the past horizon.  In
\cite{Compere:2023ktn} the constant is written as the limit of $4G H\big[S_{ij}+HP_{(i|j)}\big]$. The first term $S_{ij}$ vanishes
under the assumption made there that $T_{ij}$ is finite at the past horizon. This point is  overlooked in \cite{Compere:2023ktn, Harsh:2026vmh}.

The relation $\chi_{ij}=\eta D_-\Xi_{ij}$ ties the $1/r$ part of the field to its value on $\mathscr I^+$. In the
variables \eqref{eq:B-dictionary}, at fixed $\vec x$,
\be
\eta D_-=\eta\,\pd_\eta-1=-\frac{1}{H(1+Hr)}\,\pd_t-r\,\pd_r-1 .
\label{eq:B-Darboux}
\ee
Every multipole of $\Xi_{ij}$ is a polynomial in $1/r$, with coefficients that depend on $t_{\rm ret}$ and $n_i$.
Acting on $\Xi_{ij}=\Xi^{(0)}_{ij}+\Xi^{(1)}_{ij}/r+\dots$, the operator \eqref{eq:B-Darboux} gives
\be
\chi_{ij}=\left(1+\frac{1}{H^2r}\,\pd_t\right)\chi_{ij}\big|_{\mathscr I^+}+O\!\left(\frac{1}{r^2}\right),\qquad
\chi_{ij}\big|_{\mathscr I^+}=-\,\Xi^{(0)}_{ij},
\label{eq:B-one-over-r}
\ee
whatever $\Xi^{(1)}_{ij}$ is.  This is the relation between the $1/r$ term and the field on
$\mathscr I^+$ that was noted in Section~\ref{sec:can} and in \cite{Compere:2023ktn,Harsh:2026vmh}. It is a property that holds at every degree. 
Adding the degree-$q$ term of the sharp part to the tail term, the degree-$q$ term of $\chi_{ij}$ is
\begin{align}
\chi^{[q]}_{ij}=-4GH\eta\,\frac{(-1)^q}{q!}\,\pd_L\left[\frac{\aR^{-q}S_{ij|L}}{\rho}\right]+4GH^2\frac{(-1)^q}{q!}\,\pd_L\left[\int_{-\infty}^{t_{\mathrm{ret}}}\!\dd t'\,a^{-q}S_{ij|L}\right]
 \label{eq:comparison-raw-tensor}
\end{align}
Equations \eqref{eq:B-degree-q-1}, \eqref{eq:B-degree-q-2} and
\eqref{eq:comparison-raw-tensor} give the retarded solution at every degree in
closed form.

One could go further. The derivatives $\pd_L$ of
$\aR^{-q}X(t_\mathrm{ret})/\rho$ can be written out for general $q$, 
as powers of $1/r$. 
The
Cartesian moments of degree $q$ can be decomposed into their STF pieces of rank
$q, q-2, \dots$. 
The conservation laws
\eqref{eq:balance0}--\eqref{eq:balance2}, continued to degree $N$, can be used
to trade momentum moments for stress moments. Each of these steps is
mechanical, and the general-degree expressions that result are long without
being illuminating. 
We therefore refrain from writing them out.

Other truncations can be treated in the same way. Date and Harsh~\cite{Date:2026rft}
truncate the stress moments alone, $S_{ij|L}=0$ for every degree higher than $N$, and
let the conservation laws determine the moments of the energy and momentum densities.
For a source in our class these are then non-zero up to degree $N+2$ and $N+1$
respectively. Since \eqref{eq:B-degree-q-1}, \eqref{eq:B-degree-q-2} and \eqref{eq:comparison-raw-tensor} hold degree by degree for arbitrary moments,
they give the retarded solution in that truncation as well, with the sums running up to
$q=N+2$, $N+1$ and $N$ in the scalar, vector and tensor sectors. We do not pursue the
details here.

In Section~\ref{app:B1} we sum the
terms with $q=0,1,2$ and compare with the quadrupolar solution of
\cite{Compere:2023ktn}. In Section~\ref{app:B2} we add the $q=3$ terms and
compare with the octupolar solution of \cite{Harsh:2026vmh}.

\subsection{Quadrupolar truncation}
\label{app:B1}
Summing $q=0,1,2$ in \eqref{eq:B-degree-q-1} gives
\begin{align}
 \left.\widehat\chi\right|_{q\leq2}
 ={}&\frac{4G}{r}\left[
 \Qrp-\frac12H(\partial_{t}-H)\Qrp_{kk}
 +n_k\partial_{t} \Qrp_k
 +\frac12n_kn_l(\partial_{t}^2-H^2)\Qrp_{kl}\right]
 \nonumber\\
 &+\frac{4G}{r^2}\left[n_k\Qrp_k
 +\frac12(3n_kn_l-\delta_{kl})\partial_{t} \Qrp_{kl}\right]
 +\frac{2G}{r^3}(3n_kn_l-\delta_{kl})\Qrp_{kl}.
\label{eq:comparison-scalar-quad}
\end{align}
This is the CHK scalar solution. The vector solution follows directly from \eqref{eq:comparison-scalar-quad}:
\begin{equation}
 \left.\chi_{0i}\right|_{q\leq2}
 =\left[\, \left.\widehat\chi\right|_{q\leq2}\,\right]_{\Qrp_L\mapsto P_{i|L}}.
\label{eq:comparison-vector-replacement}
\end{equation}
Using the conservation equations \eqref{eq:balance1}--\eqref{eq:balance2} and the constraint $S_{(ij|k)}=0$, cf.~ \eqref{eq:constraints}, we get $\chi_{0i}$ in the CHK form:
\begin{align}
 \left.\chi_{0i}\right|_{q\leq2}
 ={}&\frac{4G}{r}\left[P_i-\frac H2(\partial_{t}-H)P_{i|kk}
 -n_kS_{ik}+\frac12n_kn_l(\partial_{t}+H)S_{kl|i}\right]
 \nonumber\\
 &+\frac{4G}{r^2}\left[n_kP_{i|k}-\frac12\partial_{t} P_{i|kk}
 +\frac32n_kn_l(S_{kl|i}+HP_{i|kl})\right] \nonumber\\
 &
 +\frac{2G}{r^3}(3n_kn_l-\delta_{kl})P_{i|kl}.
\label{eq:comparison-vector-quad}
\end{align}

The tensor sector is the sum of the sharp part and the tail part, cf.~\eqref{eq:comparison-raw-tensor}. At $q\le2$
the sharp part is \eqref{eq:comparison-scalar-quad} with $\Qrp_L\to S_{ij|L}$, and in the tail part the history
integral survives only through \eqref{eq:B-tail-0}. Adding the two, the result has the structure \eqref{eq:B-one-over-r}:
\bea
 \left.\chi_{ij}\right|_{q\le2}
 &=&\left(1+\frac{1}{H^2r}\,\pd_t\right)\chi_{ij}\Big|_{\mathscr I^+}
 +\frac{4G}{r^2}\Big[n_kS_{ij|k} 
 +\frac12\,\pd_t\big(3n_kn_lS_{ij|kl}-S_{ij|kk}\big)\Big]
 \nonumber \\ 
 && 
 +\frac{2G}{r^3}\big(3n_kn_lS_{ij|kl}-S_{ij|kk}\big),
\label{eq:comparison-tensor-quad}
\eea
with
\be
\chi_{ij}\Big|_{\mathscr I^+}
 =4GH^2\Big[-P_{(i|j)}+n_kS_{ij|k}+\frac12\,n_kn_l\,\pd_tS_{ij|kl}-\frac H2\,S_{ij|kk}\Big]+C_{ij}(-\infty).
\label{eq:comparison-scri-quad}
\ee
These are eqs.~(2.95) and (2.93) of \cite{Compere:2023ktn}, with $G$ restored and $C_{ij}(-\infty)$ their
$\chi_{ij}(-\infty)$. The tail part contributes only to the $r^0$ and $1/r$ terms; the $1/r^2$ and $1/r^3$ terms come from the sharp part alone.

\subsection{Octupolar truncation}
\label{app:B2}

At $N=3$ the moments of degree three are to be kept. The analogues of the constraints \eqref{eq:constraints} now come
from the degree-four equations and restrict the moments of degree three. Relative to Section~\ref{app:B1} the terms
$q=3$ are added in all three sectors, and the vector sector at $q\le2$ changes, because one of the inputs to
\eqref{eq:comparison-vector-quad} no longer holds. The term $q=3$ of \eqref{eq:B-degree-q-1} is
\begin{align}
 \widehat\chi^{[3]}
 ={}&\frac{2G}{r}\Big[\frac13\big(\pd_t^3-4H^2\pd_t\big)\,n_kn_ln_m\Qrp_{klm}
 -H\big(\pd_t^2-2H\pd_t\big)\,n_k\Qrp_{kll}\Big]
 \nonumber\\
 &+\frac{2G}{r^2}\Big[\big(2\pd_t^2-3H^2\big)\,n_kn_ln_m\Qrp_{klm}
 -\big(\pd_t^2+H\pd_t-3H^2\big)\,n_k\Qrp_{kll}\Big]
 \nonumber\\
 &+\frac{2G}{r^3}\,\pd_t\big(5\,n_kn_ln_m\Qrp_{klm}-3\,n_k\Qrp_{kll}\big)
 +\frac{2G}{r^4}\big(5\,n_kn_ln_m\Qrp_{klm}-3\,n_k\Qrp_{kll}\big).
\label{eq:comparison-scalar-oct}
\end{align}
This is eq.~(4.21) of \cite{Harsh:2026vmh}.

The vector term $\chi^{[3]}_{0i}$ follows from it by $\Qrp_{klm}\to P_{i|klm}$, as the term $q\le2$ followed from \eqref{eq:comparison-scalar-quad}. The passage from \eqref{eq:comparison-vector-replacement} to \eqref{eq:comparison-vector-quad} used the constraint
$S_{(ij|k)}=0$. At $N=3$ the degree-two momentum law in \eqref{eq:balance2} gives instead
$S_{(ij|k)}=-\tfrac12(\pd_t-H)P_{(i|jk)}$, and \eqref{eq:comparison-vector-quad} acquires the additional term
\be
3G\,n_kn_l\Big[\frac1r\big(\pd_t^2-H^2\big)+\frac{3}{r^2}\big(\pd_t-H\big)\Big]P_{(i|kl)},
\label{eq:comparison-vector-correction}
\ee
which is eq.~(5.12) of \cite{Harsh:2026vmh}.

The term $q=3$ of \eqref{eq:comparison-raw-tensor} contains no history integral, since the bracket of
\eqref{eq:B-tail-q} is differentiated three times. It has again the structure \eqref{eq:B-one-over-r}:
\begin{align}
 \chi^{[3]}_{ij}
 ={}&\left(1+\frac{1}{H^2r}\,\pd_t\right)\chi^{[3]}_{ij}\Bigg|_{\mathscr I^+}
 +\frac{2G}{r^2}\,(\pd_t-H)\left[2(\pd_t+H)\,n_kn_ln_mS_{ij|klm}-(\pd_t+2H)\,n_kS_{ij|kll}\right]
 \nonumber\\
 &+\frac{2G}{r^3}\,\pd_t\big(5\,n_kn_ln_mS_{ij|klm}-3\,n_kS_{ij|kll}\big)
 +\frac{2G}{r^4}\big(5\,n_kn_ln_mS_{ij|klm}-3\,n_kS_{ij|kll}\big),
\label{eq:comparison-tensor-oct}
\end{align}
with
\be
 \chi^{[3]}_{ij}\Bigg|_{\mathscr I^+} =\frac{2GH^2}{3}\,(\pd_t-H)\Big[(\pd_t+H)\,n_kn_ln_mS_{ij|klm}-3H\,n_kS_{ij|kll}\Big].
\label{eq:comparison-scri-oct}
\ee
These are eqs.~(4.36) and (4.37) of \cite{Harsh:2026vmh}.

In the octupolar truncation the field is therefore \eqref{eq:comparison-scalar-quad} plus
\eqref{eq:comparison-scalar-oct} in the scalar sector, \eqref{eq:comparison-vector-quad} plus
\eqref{eq:comparison-vector-correction} plus the vector version of \eqref{eq:comparison-scalar-oct} in the vector
sector, and \eqref{eq:comparison-tensor-quad} plus \eqref{eq:comparison-tensor-oct} in the tensor sector. Term by
term, it is the octupolar solution of \cite{Harsh:2026vmh}. The field on $\mathscr I^+$ is
\eqref{eq:comparison-scri-quad} plus \eqref{eq:comparison-scri-oct}, and it comes entirely from the tail part.

\bibliographystyle{JHEP}
\bibliography{biblio}
	
\end{document}